\documentclass[a4paper,12pt, epsfig]{article}
\usepackage{epsfig}
\usepackage{amssymb,latexsym}
\usepackage{amsfonts}
\usepackage{amsmath}
\usepackage{graphicx}

\newskip\humongous \humongous=0pt plus 1000pt minus 1000pt

\newif\ifdtup

\catcode`\@=11

\@addtoreset{equation}{section}
\def\theequation{\thesection.\arabic{equation}}

\def\@normalsize{\@setsize\normalsize{15pt}\xiipt\@xiipt
\abovedisplayskip 14pt plus3pt minus3pt%
\belowdisplayskip \abovedisplayskip
\abovedisplayshortskip \z@ plus3pt%
\belowdisplayshortskip 7pt plus3.5pt minus0pt}

\def\small{\@setsize\small{13.6pt}\xipt\@xipt
\abovedisplayskip 13pt plus3pt minus3pt%
\belowdisplayskip \abovedisplayskip
\abovedisplayshortskip \z@ plus3pt%
\belowdisplayshortskip 7pt plus3.5pt minus0pt
\def\@listi{\parsep 4.5pt plus 2pt minus 1pt
      \itemsep \parsep
      \topsep 9pt plus 3pt minus 3pt}}

\relax

\catcode`@=12

\catcode`\@=11

\def\section{\@startsection{section}{1}{\z@}{3.5ex plus 1ex minus
    .2ex}{2.3ex plus .2ex}{\large\bf}}

\def\thesection{\arabic{section}}
\def\thesubsection{\arabic{section}.\arabic{subsection}}

\def\appendix{\setcounter{section}{0}
  \def\thesection{Appendix \Alph{section}}
  \def\thesubsection{\Alph{section}.\arabic{subsection}}
  \def\theequation{\Alph{section}.\arabic{equation}}}

\def\SymBoxes#1#2#3#4{\newdimen\un@t \un@t#3%
\raisebox{#1}{\rule{#2\un@t}{#4}\hskip-#2\un@t
\@tempdimb\un@t \advance\@tempdimb by-#4\@tempcntb#2\relax%
\@whilenum{\@tempcntb>0}\do{
\rule{#4}{\un@t}\hskip\@tempdimb \advance\@tempcntb by\m@ne}%
\hskip-#2\un@t \rule[\un@t]{#2\un@t}{#4}%
\rule[\un@t]{#4}{#4}\hskip-#4
\rule{#4}{\un@t}}\hskip-#4}                

\begin{document}


\newcommand{\dd}{\textrm{d}}

\newcommand{\beq}{\begin{equation}}
\newcommand{\eeq}{\end{equation}}
\newcommand{\bea}{\begin{eqnarray}}
\newcommand{\eea}{\end{eqnarray}}
\newcommand{\beas}{\begin{eqnarray*}}
\newcommand{\eeas}{\end{eqnarray*}}
\newcommand{\defi}{\stackrel{\rm def}{=}}
\newcommand{\non}{\nonumber}
\newcommand{\bquo}{\begin{quote}}
\newcommand{\enqu}{\end{quote}}
\def\de{\partial}
\def\Om{\ensuremath{\Omega}}
\def\Tr{ \hbox{\rm Tr}}
\def\rc{ \hbox{$r_{\rm c}$}}
\def\H{ \hbox{\rm H}}
\def\HE{ \hbox{$\rm H^{even}$}}
\def\HO{ \hbox{$\rm H^{odd}$}}
\def\HEO{ \hbox{$\rm H^{even/odd}$}}
\def\HOE{ \hbox{$\rm H^{odd/even}$}}
\def\HHO{ \hbox{$\rm H_H^{odd}$}}
\def\HHEO{ \hbox{$\rm H_H^{even/odd}$}}
\def\HHOE{ \hbox{$\rm H_H^{odd/even}$}}
\def\K{ \hbox{\rm K}}
\def\Im{ \hbox{\rm Im}}
\def\Ker{ \hbox{\rm Ker}}
\def\const{\hbox {\rm const.}}
\def\o{\over}
\def\im{\hbox{\rm Im}}
\def\re{\hbox{\rm Re}}
\def\bra{\langle}\def\ket{\rangle}
\def\Arg{\hbox {\rm Arg}}
\def\exo{\hbox {\rm exp}}
\def\diag{\hbox{\rm diag}}
\def\longvert{{\rule[-2mm]{0.1mm}{7mm}}\,}
\def\a{\alpha}
\def\b{\beta}
\def\e{\epsilon}
\def\l{\lambda}
\def\ol{{\overline{\lambda}}}
\def\ochi{{\overline{\chi}}}
\def\th{\theta}
\def\s{\sigma}
\def\oth{\overline{\theta}}
\def\ad{{\dot{\alpha}}}
\def\bd{{\dot{\beta}}}
\def\oD{\overline{D}}
\def\opsi{\overline{\psi}}
\def\dag{{}^{\dagger}}
\def\tq{{\widetilde q}}
\def\L{{\mathcal{L}}}
\def\p{{}^{\prime}}
\def\W{W}
\def\N{{\cal N}}
\def\hsp{,\hspace{.7cm}}
\def\bo{\ensuremath{\hat{b}_1}}
\def\bfo{\ensuremath{\hat{b}_4}}
\def\co{\ensuremath{\hat{c}_1}}
\def\cfo{\ensuremath{\hat{c}_4}}
\newcommand{\C}{\ensuremath{\mathbb C}}
\newcommand{\Z}{\ensuremath{\mathbb Z}}
\newcommand{\R}{\ensuremath{\mathbb R}}
\newcommand{\rp}{\ensuremath{\mathbb {RP}}}
\newcommand{\cp}{\ensuremath{\mathbb {CP}}}
\newcommand{\vac}{\ensuremath{|0\rangle}}
\newcommand{\vact}{\ensuremath{|00\rangle}                    }
\newcommand{\oc}{\ensuremath{\overline{c}}}

\newcommand{\Vol}{\textrm{Vol}}

\newcommand{\half}{\frac{1}{2}}

\def\changed#1{{\bf #1}}

\begin{titlepage}
\begin{flushright}
IFUP-TH/2010-22 \\
SISSA  58/2010/EP
\end{flushright}
\bigskip
\def\thefootnote{\fnsymbol{footnote}}

\begin{center}
{\Large {\bf
Vortices in (2+1)d Conformal Fluids
  } }
\end{center}

\bigskip
\begin{center}
{\large  Jarah EVSLIN$^1$\footnote{\texttt{jarah@df.unipi.it}}} and {\large  Chethan KRISHNAN$^2$\footnote{\texttt{krishnan@sissa.it}}}
\end{center}

\renewcommand{\thefootnote}{\arabic{footnote}}

\begin{center}
\vspace{0em}
{$^1$\em  { INFN Sezione di Pisa, Largo Pontecorvo 3, \\ Ed. C, 56127 Pisa, Italy\\and\\
Department of Physics, University of Pisa, \\ Largo Pontecorvo 3, Ed. C, 56127 Pisa, Italy\\
\vskip .4cm}}

{$^2$\em  { SISSA and INFN - Trieste Section,\\
Bonomea-265, I-34136, Trieste, Italy\\
\vskip .4cm}}

\end{center}

\vspace{1.1cm}

\noindent
\begin{center} {\bf Abstract} \end{center}

\noindent
We study isolated, stationary, axially symmetric vortex solutions in (2+1)-dimensional viscous conformal fluids. The equations describing them can be brought to the form of three coupled first order ODEs for the radial and rotational velocities and the temperature. They have a rich space of solutions characterized by the radial energy and angular momentum fluxes. We do a detailed study of 
the phases in the one-parameter family of solutions with no energy flux. This parameter is the product of the asymptotic vorticity and temperature. When it is large, the radial fluid velocity reaches the speed of light at a finite inner radius.  
When it is below a critical value, the velocity is everywhere bounded, but at the origin there is a discontinuity.  We comment on turbulence, potential gravity duals, non-viscous limits and non-relativistic limits.



\vfill

\begin{flushleft}
{\today}
\end{flushleft}
\end{titlepage}

\hfill{}

\tableofcontents

\setcounter{footnote}{0}

\section{Introduction}

Incompressible fluids in two spatial dimensions allow vortex solutions with a rotational velocity profile that goes as $1/r$. In this note we aim to generalize vortex solutions to the relativistic case of conformal fluids with and without viscosity. We write down their equations in a tractable form, discuss their solution space, and study some of their phases in detail.

What is the relativistic generalization of the $1/r$ rotational velocity?  There are two natural guesses.  If the velocity $v$ continues to behave roughly as $1/r$ in the relativistic case, it will reach the speed of light at a finite inner radius.  On the other hand if the relativistic velocity $u=v\gamma$ is roughly $1/r$, then the velocity does not reach the speed of light at any finite radius.  Interestingly, we find that both generalizations of the $1/r$ velocity profile are realized by relativistic vortices.  When the product of the asymptotic vorticity and temperature exceeds a critical value the rotational velocity reaches the speed of light at a finite radius.  Below this critical value the velocity remains bounded, although the inward radial velocity is nonzero and so discontinuous at the origin.

\subsection{Motivation}

Vortices are likely play an essential role in turbulence in {\em any} fluid. 
The picture of incompressible turbulence as composed of superpositions of vortices dates back to Richardson \cite{Rich}, 
and is supported by evidence from analytical, numerical and experimental perspectives. The energy in such a turbulent fluid flows from modes corresponding to larger distance scales to smaller distance scales, in a phenomenon famously called ``the cascade''.  In 1941 Kolomogarov demonstrated the existence of certain scaling laws in this regime \cite{Frisch}.
In incompressible (3+1)-d fluids, 
the conjectured regularity of the flows appears to be a consequence of vortex-stretching. For the incompressible Navier-Stokes equation in 3+1 dimensions, there exists a fully regular vortex solution that is stabilized by a radial velocity, called the Burgers vortex (and its generalizations). Vortex stretching is the phenomenon by which such a vortex with non-zero inward directed radial velocity can conserve volume because matter moves outward along the third dimension. These vortex tubes are believed to be the ``sinews'' of turbulence \cite{Kida}. 
For compressible fluids in 3+1 dimensions, there is numerical evidence that vortices play a similar role (see e.g., \cite{HideakiMiuraVortexStructuresInCompressibleTurbulence}).

The nature of (2+1)-dimensional vortices is quite different. 
The viscous dissipation may be arbitrarily small in 2+1 dimensions, and so one can work with the vortices of the non-dissipative (Eulerian) fluid, which have a divergent rotational velocity at the core. The statistical mechanics of these vortices (e.g.,\cite{Onsager}) is in broad agreement with experiments and numerical simulations\footnote{In working with inviscid vortices, we ignore energy and enstrophy dissipation. The latter is a questionable assumption at high Reynolds numbers, but it has been argued that this assumption does not invalidate the results \cite{Sommeria}.}. In contrast, the (3+1)-d Burgers vortex depends on the viscosity crucially for its existence\footnote{The fact that there exists a regular vortex solution that is sustained by viscosity is undoubtedly of relevance to the existence and smoothness of the solutions to the Navier-Stokes equations. Burgers vortex has non-vanishing velocity along the third axis. So contrary to what one might naively expect, the viscosity need not dissipate the flow to an equilibrium, as it would in compact domains with decaying boundary conditions. In (2+1)-dimensional fluids, the third dimension is of course unavailable, however vortices with velocities that blow up at the core are still a very useful approximation for understanding the statistical behavior of turbulence.}. Another difference in (2+1)-d incompressible turbulence 
is the appearance of an inverse energy cascade \cite{Kraichnan, mcwilliams}: 
energy is transmitted to bigger and bigger length scales as vortices coalesce.

Recently, the AdS/CFT correspondence \cite{AdSCFT} has opened up a new perspective on the old problem of fluids and turbulence. The claim is that the fluids that arise as the long-wavelength limits of (a large class of) conformal quantum field theories are expected to be described by gravity through the fluctuations of black objects in an asymptotically anti-de Sitter space, as described in Refs.~\cite{andrei1, andrei2, Min1, Min2, Strominger}. These references provide a map from reasonably generic fluid configurations to solutions of Einstein's equations up to some order in derivatives.  Needless to say, if this implies that turbulence is as ubiquitous in gravity as in fluid mechanics, it would be difficult to overstate its relevance for example for cosmology.  Indeed the chaotic evolution of gravity near a big bang singularity in the BKL ansatz is already well established \cite{BKL}.  On the other hand it is unclear at the moment how useful this map is going to be for a direct understanding of turbulence, but the fact that Einstein's gravity can be dual to fluid dynamics is interesting in itself, and worthy of investigation. Therefore, because of the crucial importance of vortices in fluid dynamics, in this paper we initiate a study of vortex solutions in relativistic conformal fluids.

One motivation for studying conformal fluids comes from the heavy ion collisions at RHIC and the LHC. It seems possible that the state of matter that results from these collisions is best described by a strongly coupled phase of QCD plasma. This plasma is well-described by a turbulent \cite{Romatschke} conformal fluid with a very small viscosity $\eta$. AdS/CFT methods have been applied to study this system with very suggestive, but still only semi-quantitative, success \cite{RHIC}. The onset of turbulence is expected to happen at high Reynolds numbers and for a relativistic fluid at temperature $T$, this is determined by
\bea
{\rm Re} \sim \frac{s T L}{\eta} \gg 1.
\eea
Here $s$ is the entropy density and $\eta$ is the viscosity, and there are experimental and theoretical reasons to believe that $\eta/s$ is close to its conjectured lower bound, namely $1/4\pi$. For the RHIC plasma the relevant length scale $L$ is the size of the gold nucleus, and the temperature is the QCD scale. Plugging in these numbers, one finds that ${\rm  Re} \sim 150 \gg 1$. This suggests at least naively, that one should expect that the plasma is turbulent, and a first step towards an understanding of this deeply non-linear regime is the study of vortices. In heavy ion collisions, the standard assumption is that one is dealing with a boost-invariant plasma in the longitudinal directions. This is the so-called ``Bjorken flow'' \cite{Bjorken} and it makes the relevant fluid dynamics two (spatial) dimensional. This is one of the reasons why we focus on vortices in 2+1 dimensions. The (3+1)-dimensional case is also certainly of interest: in particular, it will be interesting to see if there are fully regular vortex solutions (like the Burgers vortex in incompressible fluids) in conformal fluids. See \cite{Vishnu} for a different approach to turbulence via string theory and holography.

Consequences of the potential existence of vortices in neutron stars have also received some attention \cite{Savvidy}.  However the vortices of interest contain (color) magnetic flux, and so the purely hydrodynamic vortices of the present note may be too primitive to shed light on their rich phenomenology.  Vortices in neutron stars are generally treated in the superfluid approximation, however viscosity is necessary in a large region of the interior of a neutron star to avoid an $r$-mode instability that would be in conflict with the observed rotational velocities \cite{Rajagopal}. This provides one motivation for our interest in vortices in viscous fluids.

Apart from their possible appearance in turbulent flows, vortices might also be relevant for an understanding of gravity solutions in AdS. Recently certain black funnel solutions which have horizons extending to the boundary of AdS were constructed \cite{MarolfRang, Hubeny2} using AdS C-metrics \cite{Plebanski}. They were interpreted on the boundary as a fluid in equilibrium with a boundary black hole, possibly providing a holographic construction of Hartle-Hawking vacua for strongly coupled field theories\footnote{The C-metric based approach of \cite{MarolfRang} does not work for constructing geometries where the boundary is flat, so these were done for geometries with hyperbolic spaces or compact spaces on the boundary.}. A natural generalization would be to include rotation along an axis on the boundary. This corresponds to a spinning black funnel that has a horizon stretching all the way to the boundary. Of course, the spin has to be such that one does not run into trouble with superluminal rotation and similar pathologies. Constructing such solutions using generalized AdS C-metrics seems possible, see also \cite{EH}. The relation between fluids and gravity have generalizations away from AdS as well. The membrane paradigm for black holes suggests that black hole horizons can be treated in terms of fluid dynamics \cite{Damour, Thorne}. There has been recent work on the construction of gravity solutions using the blackfold approach \cite{blackfold} (generalizing previous work on the construction of exact vacuum solutions in higher dimensions \cite{5d}). In AdS, the relation between the boundary fluid picture and the horizon fluid picture was connected via a Wilsonian RG flow in \cite{Strominger}.

In fact a large class of the solutions that we find in this paper have a core of finite radius where the fluid quantities diverge, while asymptotically they tend to those of a static black brane. This structure has similarities to what could be the dual of a spinning black funnel-like solution that might exist in AdS. We emphasize, however, that at this stage this is only an analogy.

In this paper we survey vortices in a conformal relativistic fluid in flat (2+1)-dimensional Minkowski spacetime. The appropriate generalization of this construction to other (2+1)-dimensional geometries which are likely to have gravitational duals is left for the future. An explicit gravity dual will enable us to construct the boundary stress tensor directly and see how it compares with the fluid stress tensor.

\subsection{Summary of Results}

We numerically find axially symmetric vortex solutions which asymptote to a static fluid configuration. The latter condition means that we are studying isolated vortices, deferring the harder problem of vortex interactions. 
Correspondingly, from the perspective of the dual black objects in AdS, we expect the geometry to asymptote to a  static AdS black brane. This should be dual to a static fluid configuration at constant temperature in the asymptotic region of the  {\em boundary}. Indeed, in the black funnel cases that have been investigated, it is known that the fluid does tend to a static ideal fluid configuration in this asymptotic region \cite{Hubeny2}. In any event, the boundary conditions that we impose on our solutions are that the velocities should fall off at infinity and that the temperature should have a leading behavior that is constant. It should be noted that implicit in this assumption is the condition that the solutions are well-defined for large $r$.


Our concern will mostly be viscous conformal fluids in this note, even though we also comment on the ideal limit. The equations become significantly more complicated when we include viscosity. One of the results of this paper is that these equations can be brought to the form of three coupled first order ordinary differential equations, one each for the temperature, the radial velocity and the rotational velocity (see Appendix B). In order to make statements about the asymptotic behavior, we will assume that the velocities and temperature can be expanded in inverse powers of $r$ in the asymptotic region. This is a sufficient condition for asymptotic staticity, but is not strictly necessary.  
We find that there are many classes of such solutions that one can find, and that all of them that are asymptotically static have an inward radial velocity. In fact, this is not too surprising because in viscous fluids rotation causes an outward radial energy-momentum transfer between the fluid layers, which may be balanced by an inward radial mass flow. 
If one allows no radial flux of matter-energy or angular momentum in the vortex, then one finds that there are no asymptotically static solutions that are allowed by the fluid equations. If on the other hand, one allows such fluxes, then solutions are possible. We study one case in some detail, one where angular momentum flows radially from infinity to a sink at the center of the vortex. It exhibits two phases, both of which have some form of singular behavior at the core, as in the incompressible case. This singularity can be at finite radius or at the center of the vortex, depending on the parameters.  The rotational velocity diverges\footnote{More precisely, the coordinate velocity diverges, while the proper velocity tends to the speed of light.} at the core in the former phase. For vortices in the latter phase, the rotational velocity is finite everywhere, but the radial velocity is non-zero at the origin resulting in a discontinuity. The temperature diverges at the core in both phases. The radial component of the velocity (we always work in Landau gauge) is always nonzero and in fact is always inwards.  This is analogous to the Burgers vortex, but of course without the extra dimension.

Conformal fluids at rest have only one scale, set by the temperature.  The other quantities, such as the pressure, density and viscosity, are proportional to powers of the temperature.  The constants of proportionality that we will use for numerics are those that arise from the correspondence to AdS black branes, however we will see that any other choice of coefficients simply leads to certain rescalings of our solutions.

The hydrodynamics of a neutral relativistic fluid is completely characterized by the conservation of a stress tensor.  In our axially-symmetric, stationary ansatz this consists of three ordinary differential equations, representing the local conservation of the components of the momentum 3-vector.
In the case of the viscous, conformal fluids of interest, these equations involve the temperature and its first derivative as well as the rotational and angular velocities and their first two derivatives.  Therefore the solutions are characterized by five constants of integration.  Fixing an overall scale and specializing to the case with zero radial energy flux, one is left with 3 dimensionless parameters.  We will find that only a 1-parameter family of these satisfies our asymptotically static condition.

More explicitly, these conservation equations can be brought to the form of three first order equations: one for the radial velocity, one for the rotational velocity and one for the temperature. These equations explicitly involve two constants of integration, $c_1$ and $c_2$, which we will relate to the radial energy and angular momentum flux respectively.  Clearly the counting of solutions of these equations must proceed as outlined above.  In units where the velocity of light and the Newton's constant in the gravity dual are set to unity (cf. (\ref{P1}-\ref{P2})), the vortex solutions are therefore determined by five dimensionful quantities, or four dimensionless ones\footnote{Note that the transport coefficients appearing in the equations are part of the description of the system, not the solutions.}.
The case with no energy flux from infinity corresponds to setting $c_1=0$, and in this case, the asymptotic static constraint kills two of these. The remaining dimensionful scales can be taken to be the asymptotic temperature $T_0$ and the vorticity $\omega$.  When we speak of vorticity, we will mean the integrated nonrelativistic vorticity, which is the integral of the rotational velocity around the vortex on a contour far enough away that the velocity is much less than the speed of light.  By rotational velocity we mean the velocity in the angular direction, which has dimensions of velocity and is equal to the angular velocity times the radius.

The only dimensionless quantity characterizing the ($c_1=0$) vortex is the product $\omega T_0$.  The phase of the vortex and the velocity at its core may only depend on this combination.  We will see that when $\omega T_0\gtrsim 0.1$, the rotational velocity reaches the speed of light at a finite radius, which we will call $\rc$.  This radius therefore must be equal to $\omega$ times a function of $\omega T_0$.  The (relativistic) radial velocity at $\rc$ we will see is always inwards, and is equal to $\sqrt{3}/2$.  The true radial velocity, which is the relativistic radial velocity divided by $\gamma$, therefore goes to zero at $\rc$.  By velocity we will always mean the relativistic velocity unless specified otherwise.  The rotational velocity diverges like $(r-\rc)^{-1/2}$.  On the other hand when $\omega T_0\lesssim 0.1$, the velocity never reaches the speed of light.  The rotational velocity decreases to zero at the origin, while the inward radial velocity tends to a constant which varies from 0 when $\omega T_0=0$ to $1/\sqrt{2}$ when $\omega T_0\sim 0.1$.  The temperature at the core however diverges like $1/r$. 

The divergences in the core are clearly unphysical. In the case of an incompressible fluid, in a physical flow they are always avoided. However the $1/r$ vortex solution is still useful because it serves as an idealization in terms of which the statistical properties of the flow can be described. Such divergences are avoided in real situations because it is impossible to accelerate a fluid particle to the velocity of light during a finite time $t$.  Our solutions should instead be interpreted as the universal $t\rightarrow\infty$ limits of vortices subjected to certain boundary conditions for a time $t$.  At all finite times $t$ the solutions are nonsingular.

This argument also applies to conformal fluids. It should however be mentioned that in extreme regimes where the various derivatives are large compared to the temperature of the fluid, we should in principle include higher order transport coefficients in the fluid equations to determine the correct flow \cite{IS}. But whether the flow goes to a regime where higher derivative effects become relevant depends on the initial conditions of the flow: a million dollars are available to those who solve a very closely related problem in a very closely related situation. This is not our concern here, we are interested in stationary configurations which can function as building blocks of a more complicated flow.

In the next section, as a simple warmup we write down the most general axially symmetric vortex solution for incompressible Navier-Stokes equations in 2+1 dimensions. The $1/r$ vortices that have been used in the statistical study of (2+1)-dimensional turbulence are a special class of these. Then we turn to a description of conformal fluid dynamics and the equations that describe stationary vortices in them. In the remaining sections,  we go on to study various kinds of vortices from various angles. We consider different phases of ideal and dissipative fluids analytically determining their limiting behavior far from and near to the core and then numerically finding the entire solutions.  While a quantitative understanding of the role played by vortices in (conformal) turbulence is lacking at present, we believe that an exploration of the space of these quasi-solitonic solutions is the place to start.

\section{Incompressible (2+1)-dimensional Fluids} \label{incsec}

In the absence of forcing, incompressible flows in (2+1)-dimensions solve the Navier-Stokes equation
\bea
\partial_t \xi + {\bf v}\cdot\nabla \xi=\nu \nabla^2 \xi, \ \ \nabla \cdot {\bf v}=0
\eea
where ${\bf v}=(v_x,v_y)$ is the velocity, $\xi=\partial_x v_y-\partial_y v_x$ is the vorticity and $\nu$ is the kinematical viscosity. We will search for axially-symmetric vortex solutions of these equations in which the velocity asymptotes to a constant at radial infinity, corresponding to the ansatz
\bea
(v_x, v_y)=(-y f(r) + x g(r), xf(r)+y g(r) ). \label{inans}
\eea
In this section the functions $f(r)$ and $g(r)$ will stand for non-relativistic quantities, everywhere else we will consider their appropriate relativistic generalizations.
The incompressibility immediately forces a condition on the expansion function $g$:
\bea
g(r)=\frac{k}{r^2}, \label{gin}
\eea
where $k$ is an integration constant. Substituting this back into the ansatz and using the vorticity equation of motion, we find a simple differential equation for $f$ whose general solution is
\bea
f(r)=\frac{C_1}{r^2}+C_2+C_3 r^{k/\nu}.\label{fin}
\eea
Equations (\ref{fin}) and (\ref{gin}) together comprise the general solution of the stationary unforced axially symmetric equations in 2+1 dimensions\footnote{\label{incomp}There are some caveats to this statement. One is when there is no radial flow, $k=0$. Then the most general solution is of the form
\bea
f(r)=\frac{C_1}{r^2}+C_2+C_3 \log r,\label{fin2}
\eea
for any non-zero viscosity $\nu~$. When there is no viscosity, but the radial velocity is non-zero ($k\neq 0$), then the general solution takes the form
\bea
f(r)=\frac{C_1}{r^2}+C_2.
\eea
When viscosity and radial velocity are both zero, {\em any} function $f(r)$ is a solution.
}. Note that to obtain the actual radial and angular velocity, one must multiply $g$ and $f$ by $r$.

It is evident that there may be solutions that fall-off at infinity if one sets $C_2=0$, so long as the radial flow is either zero or inward (in the latter case $k$ is negative and should satisfy $|k|>\nu$). In a non-relativistic theory, the function $g$ directly captures the radial inward flux:  we can integrate the radial velocity $ r g(r)$ around any circle and we get a flux $\sim k$. The statistical mechanics of the $f \sim 1/r^2$ vortices is a useful model for two-dimensional turbulence \cite{Kraichnan, Onsager}.

\section{Equations of Motion for Vortices in Conformal Fluids}\label{eom}

In the rest of this note we will generalize the discussion of Sec.~\ref{incsec} to the case of stationary, relativistic, axially-symmetric vortices in a (2+1)-dimensional conformal fluid. Conformality implies that all of the transport coefficients of the fluid are equal to constants times powers of the temperature $T$. To determine these coefficients unequivocally, we need to have full control over the underlying CFT or the gravity dual.  However we will see that changes in the values of these constants can easily be reabsorbed into a rescaling of the solutions.   Thus, we lose no generality by choosing these constants to be those arising from the black brane ansatz considered in \cite{Min2}. 

Because of axial symmetry, every object must transform covariantly under the SO(2) axial symmetry. The spatial velocity is an SO(2) vector.  Therefore it must be of the form
\beq
u^x=-yf(r)+xg(r)\hsp u^y=xf(r)+yg(r)
\eeq
where $u^x$ and $u^y$ are the two spatial components of the relativistic 3-velocity, in other words, they are the true velocity multiplied by
\beq
\gamma(r)=\sqrt{1+(u^x)^2+(u^y)^2}=\sqrt{1+r^2(f^2(r)+g^2(r))}.
\eeq
We will use the word ``velocity'' for $u$.  $f(r)$ and $g(r)$ are arbitrary functions of the radial coordinate.  $f(r)$ is the angular velocity, $rf(r)$ is the rotational velocity and $rg(r)$ is the radial velocity.  $u$ is a function of $x$ and $y$, but the dependence will be left implicit.  We will also sometimes omit the $r$ dependence of $f,$\ $g$ and $\gamma$.  Indices will be raised using the metric $\eta^{\mu\nu}$ with signature $(-1,1,1)$.  In particular this yields the temporal component of the velocity
\beq
u^0=\gamma(r)=-u_0
\eeq
and, being timelike
\beq
u^\mu u_\mu=-1.
\eeq

There is no canonical definition of the velocity $u$ of a fluid.  We will use the velocity of the Landau frame, which is the frame in which the stress tensor consists of an isotropic non-dissipative term plus a dissipative term which is entirely spacelike.  One may then obtain the spatial components in this frame by contracting with the projection
\bea
\mathcal{P}^{\mu\nu}&=&\eta^{\mu\nu}+u^\mu u^\nu\\&=&\left(
\begin{array}{ccc}
r^2(f^2+g^2)&(-yf+xg)\gamma&(xf+yg)\gamma\\
(-yf+xg)\gamma&1+y^2f^2+x^2g^2-2xyfg&xy(g^2-f^2)+(x^2-y^2)fg\\
(xf+yg)\gamma&xy(g^2-f^2)+(x^2-y^2)fg&1+x^2f^2+y^2g^2+2xyfg
\end{array}
\right).\nonumber
\eea
Being a projector, all eigenvalues of $\mathcal{P}$ are equal to zero or one.  By construction the unique zero-eigenvector is the velocity $u$
\beq
\mathcal{P}^{\mu\nu}u_\nu=\eta^{\mu\nu}u_\nu+u^\mu u^\nu u_\nu=u^\mu-u^\mu=0.
\eeq
Also it is easy to see the $\mathcal{P}$ squares to itself.

The dynamics of an uncharged relativistic fluid is simply described by the conservation of the stress tensor, which to first order in a derivative expansion is
\beq
T^{\mu\nu}=\rho u^\mu u^\nu+P\mathcal{P}^{\mu\nu}-\eta\sigma^{\mu\nu}-\zeta\theta\mathcal{P}^{\mu\nu}. \label{T}
\eeq
Naturally the derivative expansion is only reliable only as long as the derivatives are small compared to the temperature.



$P$ is the pressure.  In a conformal theory the stress tensor must be traceless.  We will define the pressure and density so that these first two terms, which are those of an ideal fluid, are traceless without the first derivative terms.  Therefore
\beq
\rho=2P.
\eeq
The third and fourth terms of (\ref{T}) are proportional to derivatives of the velocity.  $\eta$ is the shear viscosity, and $\zeta$ the bulk viscosity.  $\theta$ is the 3-divergence of the velocity, and $\sigma$ is a traceless symmetric tensor equal to the shear, which will be defined momentarily.  The quantities $\rho$, $P$, $\eta$ and $\zeta$ are all scalars, and so by axial symmetry can only depend on the radial coordinate $r$.
The entire stress tensor must be traceless, and so we must also impose the tracelessness of the first order derivative terms.  This implies that the bulk viscosity $\zeta$ vanishes and the stress tensor may be simplified to
\beq
T^{\mu\nu}=P(\eta^{\mu\nu}+3u^\mu u^\nu)-\eta\sigma^{\mu\nu}. \label{T2}
\eeq

The matrix $\sigma$ is the shear, which is the spatial, traceless, symmetric part of the first derivative of the velocity.  To make it spatial, one need only multiply by the projector $\mathcal{P}$, and to make it traceless, one subtracts the trace.  One may subtract the trace before the projection, and it remains traceless because the trace of the terms projected out are of the form
\beq
u^\mu\partial_\nu u_\mu=\frac{1}{2}\partial_\nu (u^\mu u_\mu)=\frac{1}{2}\partial_v(-1)=0.
\eeq
Therefore the matrix $\sigma$ is
\beq
\sigma^{\mu\nu}=\mathcal{P}^{\mu\alpha}\mathcal{P}^{\nu\beta}(\partial_\alpha u_\beta+\partial_\beta u_\alpha -\eta_{\alpha\beta}
\partial_\rho u^\rho)
\eeq
where we have used the fact that there are two spatial dimensions in the normalization of the divergence term.

The component $\sigma^{tt}$ is an SO(2) scalar, the components $\sigma^{tk}$ form a vector and $\sigma^{jk}$ is a symmetric 2-tensor.  Therefore they may be written
\bea
&&\sigma^{tt}=\sigma^{tt}(r)\hsp
\sigma^{tx}=\frac{x}{\gamma}A(r)+\frac{y}{\gamma}B(r)\hsp
\sigma^{ty}=\frac{y}{\gamma}A(r)-\frac{x}{\gamma}B(r)\nonumber\\
&&\sigma^{xx}=x^2C(r)+y^2D(r)+xyE(r)\hsp
\sigma^{xy}=(y^2-x^2)\frac{E(r)}{2}+xy(C(r)-D(r))\nonumber\\
&&\sigma^{yy}=x^2D(r)+y^2C(r)-xyE(r) \label{sig}
\eea
where as always $\gamma$ depends on $r$, but the dependence is left implicit, as will be the dependences of $A$, $B$, $C$, $D$ and $E$.  A few pencils later one can express these functions of $r$ in terms of $f$ and $g$
\bea
\sigma^{tt}&=&2r^3fgf\p+r^3(g^2-f^2)g\p\\
A&=&-r^2f^2(f^2+g^2)+r(1+r^2g^2)(ff\p+gg\p) \label{AA}\\
B&=&-r^2(f^2+g^2)fg-rgf\p(1+2r^2f^2+r^2g^2)+rfg\p(1+r^2f^2)\\
C&=&-2f^2g+\frac{1}{r}(1+r^2g^2)g\p \label{CC}\\
D&=&2f^2g+2rfgf\p-\frac{1}{r}(1+r^2f^2)g\p \label{DD}\\
E&=&2f^3-2fg^2-\frac{2}{r}(1+r^2g^2)f\p. \label{EE}
\eea
In the stationary  ansatz $\sigma^{tt}$ and $B$ do not affect the equations of motion.  Notice that, as we have set the speed of light to unity, velocity is dimensionless and so $f$ and $g$ have the same dimensions as $1/r$.  The shear, $\sigma$, is the derivative of a velocity and so it also has dimensions $1/r$, which implies that $A$ and $B$ have dimensions $1/r^2$ and $C$, $D$ and $E$ have dimensions $1/r^3$.

So far our discussion has been applicable to any conformal fluid.  Any conformal fluid has a single scale, the temperature $T$.  All other thermodynamic quantities may be expressed as monomials in $T$.  In (2+1)-dimensions, $\eta$ is proportional to $T^2$ and $P$ to $T^3$.
\beq
\eta=a T^2\hsp
P=b T^3. \label{P1}
\eeq
In fluid flow dual to a (perturbed) black brane in AdS, these constants of proportionality are fixed in terms of the Newton's constant $G$ of the gravitational theory \cite{Min2}.
\bea
a=\frac{\pi}{9G}\hsp b=\frac{4\pi^2}{27G}. \label{P2}
\eea
In fact, our numerical results are easily generalizable to the more general conformal fluid.  In a generic conformal fluid, the definitions of $\eta$ and $P$ in terms of $T$ may be multiplied by some numerical constants $C_1$ and $C_2$ respectively. We will see that if $C_1=C_2$ then the equations of motion (\ref{alpha},\ref{calc},\ref{cald},\ref{cale}) for the velocities are invariant, although both constants of integration ($c_1$ and $c_2$, cf. (\ref{stokes2}), (\ref{ang})) must be multiplied by $C_1$.  For general values of $C_1$ and $C_2$, one obtains the same equations but with the temperature $T$ multiplied by $C_1/C_2$ and the constants of integration multiplied by $C_1^3/C_2^2$.   Therefore the velocity profiles that we find for general values of $c_1, c_2$ and $T$ can be straightforwardly translated to the general case.

Putting together Eqs.~(\ref{T2}, \ref{sig}, \ref{AA}, \ref{CC}, \ref{DD}, \ref{EE}, \ref{P1}, \ref{P2}) we can express the stress tensor as a function of $f$, $g$ and $T$.  It will be convenient to decompose the stress tensor into $SO(2)$ tensors, as we did for $\sigma$
\bea
&&T^{tt}=T^{tt}(r)\hsp
T^{tx}=x\mathcal{A}(r)+y\mathcal{B}(r)\hsp
T^{ty}=y\mathcal{A}(r)-x\mathcal{B}(r)\nonumber\\
&&T^{xx}=x^2\mathcal{C}(r)+y^2\mathcal{D}(r)+xy\mathcal{E}(r)\hsp
T^{xy}=(y^2-x^2)\frac{\mathcal{E}(r)}{2}+xy(\mathcal{C}(r)-\mathcal{D}(r))\nonumber\\
&&\sigma^{yy}=x^2\mathcal{D}(r)+y^2\mathcal{C}(r)-xy\mathcal{E}(r) \label{Tabcde}
\eea
and write the equations of motion directly in terms of the functions of the radius that appear in those tensors.  The equations of motion are just the conservation of the stress tensor.  As the configuration is stationary, the time derivatives vanish.

Consider first the conservation of energy
\beq
0=\partial_x T^{xt}+\partial_y T^{yt}=2\mathcal{A}+r\mathcal{A}\p. \label{stokes1}
\eeq
This equation may be integrated to solve for $\mathcal{A}$, yielding
\beq
\mathcal{A}=\frac{c_1}{r^2} \label{stokes2}
\eeq
where $c_1$ is a constant of integration. 
We will later interpret $c_1$ as the radial energy flux, which falls into a sink at the core of the vortex when $c_1\neq 0$. This is easily seen by remembering the definition of the stress tensor as the flux of the $\mu$-th component of momentum 3-vector through a surface at fixed $\nu$ coordinate. The energy flux is computed as
\bea
{\rm Flux}=\int_\Sigma T^{t \nu} d\Sigma_\nu \sim \int_{\Sigma_r} (T^{t x} dy -T^{t y} dx) \sim \int_{r=const} r^2 \mathcal{A} \ d\theta = 2 \pi c_1.
\eea
Because of the local conservation law we can move the integral to a circle of constant radius, which we denote $\Sigma_r$. We have changed from Cartesian to polar coordinates using $x=r \cos \theta, \ y= r \sin \theta$, while noticing that the surface element (which is one-dimensional in two spatial dimensions) goes as $(dy, -dx)$. The final result arises from Stokes' theorem applied to (\ref{stokes1}, \ref{stokes2}) if there is a source/sink at the origin with strength $\sim c_1$. We will mostly be interested in vortices with no net energy flux, so we will set $c_1=0$ in most of this paper.

Next we will impose the conservation of momentum in the $x$ direction
\beq
0=\partial_x T^{xx}+\partial_y T^{yx}=x(3\mathcal{C}+r\mathcal{C}\p-\mathcal{D})+\frac{y}{2}(4\mathcal{E}+r\mathcal{E}\p). \label{px}
\eeq
This equation must be satisfied at every point in space, therefore the coefficients of $x$ and $y$ must vanish separately.  This yields two equations
\beq
\mathcal{D}=3\mathcal{C}+r\mathcal{C}\p
\eeq
and
\beq
4\mathcal{E}+r\mathcal{E}\p=0.
\eeq
The second equation may be integrated to solve for $\mathcal{E}$, yielding
\beq
\mathcal{E}=\frac{c_2}{r^4} \label{ang}
\eeq
where $c_2$ is a constant of integration which we will see is proportional to the product of the asymptotic viscosity and angular velocity.  The equation for conservation of momentum in the $y$ direction is related to (\ref{px}) by an SO(2) transformation, so it can give no new constraints on the SO(2)-invariant functions. An analogous computation to the one we did for the energy net-flux can be done for momentum as well. In polar coordinates this will become angular momentum, with the resulting condition that $c_2=0$ should hold for the net angular momentum to  be zero.

The functions $\mathcal{A},\ \mathcal{B},\ \mathcal{C},\ \mathcal{D},$ and $\mathcal{E}$ may be expressed in terms of the coefficients (\ref{AA},\ref{CC},\ref{DD},\ref{EE}) by substituting their definition Eq.(\ref{sig}) into Eq.~(\ref{T}).  We will see that $\mathcal{B}$ does not appear in the equations of motion, the others are equal to
\bea
&&T^{tt}=(2+3r^2(f^2+g^2))P-\eta\sigma^{tt}\hsp
\mathcal{A}=3\gamma g P-\frac{\eta}{\gamma}A\\
&&\mathcal{C}=(\frac{1}{r^2}+3g^2)P-\eta C\hsp
\mathcal{D}=(\frac{1}{r^2}+3f^2)P-\eta D\hsp
\mathcal{E}=-6fgP-\eta E.\nonumber
\eea
Thus we may summarize by writing a complete set of equations of motion
\bea
3\gamma g P-\frac{\eta}{\gamma}A&=&\frac{c_1}{r^2}\label{alpha}\\
\mathcal{C}&=&(\frac{1}{r^2}+3g^2)P-\eta C\label{calc}\\
(\frac{1}{r^2}+3f^2)P-\eta D&=&3\mathcal{C}+r\mathcal{C}\p\label{cald}\\
6fgP+\eta E&=&-\frac{c_2}{r^4}\label{cale}
\eea
where $P$ and $\eta$ are given in Eq.~(\ref{P1}, \ref{P2}) in terms of the temperature and the Newton's constant in the gravitational theory.  Eq.~(\ref{alpha}) is the conservation of energy, Eqs.~(\ref{calc}) and (\ref{cald}) are the conservation of momentum in the radial direction and Eq.~(\ref{cale}) is the conservation of angular momentum.  The equations of motion are all proportional to the inverse Newton's constant. Therefore Newton's constant may be multiplied out of the equations of motion, and so does not affect the equations for the velocities and temperature although it provides a multiplicative constant in the viscosity and pressure.

\section{Ideal Conformal Vortices}\label{ideal}
We will start by looking for vortex solutions in ideal fluids. The ideal fluid flow equations are obtained by setting $\eta=0$ in the equations of the previous section. The equations that we need to solve take the form
\bea
3 \gamma g P-\frac{c_1}{r^2}=0, \ \ \ 6 f g P+\frac{c_2}{r^4}=0 \hspace{0.7in} \\
r\Big[\Big(\frac{1}{r^2}+3 g^2\Big)P \Big]' + 3 \Big[\Big(\frac{1}{r^2}+3 g^2\Big)P \Big]=\Big(\frac{1}{r^2}+3 f^2\Big)P
\eea
An asymptotic expansion of the form (see discussion in section \ref{5.1})
\bea
f&=&
\frac{f_ 2}{r^2}+\frac{f_3}{r^3}+... \\  \label{asp1}
g&=&
\frac{g_2}{r^2}+\frac{g_3}{r^3}+... \\
P&=&P_0+\frac{P_1}{r}+\frac{P_2}{r^2}+\frac{P_3}{r^3}+... \label{asp2}
\eea
yields the constraints
\bea
c_1=3 g_2 P_0, \ c_2=-6 f_2 g_2 P_0,
\eea
on $f_2$ and $g_2$ with no constraints on $P_0$. The higher order coefficients are determined recursively from these. This suggests that there are indeed asymptotically static vortex solutions in the non-viscous case.

An interesting fact is that in the ideal case, these solutions can be analytically determined. The way to do this is to solve the first two equations (which are algebraic) explicitly and then solve the remaining first order ODE directly. The solutions are of the form
\bea
f(r)=\pm \sqrt{{c_2^2 a^2 \over 2 r^4} - {2 c_1^2 a^2 \over r^2} \pm
    {(4 c_2^2 a^2 r^6 + (-c_2^2 a^2 r^2 +
          4 c_1^2 a^2 r^4)^2)^{1/2} \over 2r^6}},
\eea
where $a$ is  an integration constant and there is no correlation between the signs. Once $f$ is known, $g$ and $P$ can be determined algebraically via the constraint equations. At first sight it might seem that these solutions do not allow asymptotically static solutions, but one should remember that the integration constant can be imaginary as long as the final solution is real. Indeed it is possible to show that by choosing $a$ appropriately, solutions that satisfy the asymptotic expansion can be found.

When $c_1$ and $c_2$ are both zero and we set $g=0$, we can also  see a different branch of solutions: any function $f(r)$  is a solution with the temperature (and therefore pressure) determined by
\bea
T(r)=T_0 \exp \Big(\int_{r_0}^r f(r)^2 r dr\Big). \label{nonvisc}
\eea
This function-worth of solutions is analogous to the solutions we found in the incompressible case, see footnote \ref{incomp}.

Now we turn to the discussion of vortex solutions in the presence of viscosity for generic values of $c_1$ and $c_2$. 


\section{Asymptotics and Phases of Viscous Conformal Vortices} \label{assec}

In this section we consider vortices in dissipative conformal fluids by including the viscosity term and consider the full equations of motion that we found in Section \ref{eom}. These equations consist of a first order equation (\ref{cale}), a second order equation (using (\ref{calc}) as definition of ${\cal C}$ in (\ref{cald})) and a constraint that we will use to obtain the temperature\footnote{The transport coefficients of a conformal fluid are powers of the temperature times a constant.} (\ref{alpha}). There are various kinds of non-linearities in these equations. An analytic solution is likely to be difficult to find, so we will try to find numerical solutions of this system. The trouble is that Mathematica fairs better with differential equations that are of the same order, rather than with a mixed system especially when there are constraints. So as a first step we will find a transformation that brings these equations to a set of three first order equations of the structurally very simple form
\bea
f'(r)=F(f,g,T,r), \ \ g'(r)=G(f,g,T,r), \ \ T'(r)=H(f,g,T,r),
\eea
for known (but somewhat complicated) functions $F, G$  and $H$. The details and some explicit forms are given in Appendix B. 

We will seek solutions of these equations that are ``asymptotically ideal''. That is, we want them to go to a static fluid with constant non-zero temperature far away from the core. 
We discuss some features of the $c_1\neq 0 \neq c_2$ case first and then turn to a rather detailed analysis of the $c_1=0$ case in the following subsections. Finally we comment on the case where $c_1=c_2=0$ and discuss some subtleties involved in this case. Our aim is only to construct {\em some} solutions and not attempt an exhaustive scan of the solution-space.

When $c_1=0$ there is no energy flux from infinity into a sink at the origin and when $c_2=0$ there is no such angular momentum flux. Of course, in a realworld fluid flow singular sinks do not exist, but they are useful idealizations that capture dissipative features of the flow. In the incompressible (non-relativistic) case, if there is no radial flow there are no sinks: this is because material transfer is necessary for flux of either kind to exist.
The situation is more subtle in the relativistic case we consider. The condition for no flux is not simply the vanishing of $g$: the precise conditions  for no flux are given by the vanishing of $c_1$ and $c_2$ as we showed in the last section. In fact, we will see that asymptotically static vortices with both $c_1$ and $c_2$ zero are impossible in dissipative conformal fluids (unlike in incompressible fluids).


\subsection{A General Class of Vortices\label{5.1}}

\begin{figure}
\begin{center}
\includegraphics[width=8cm]{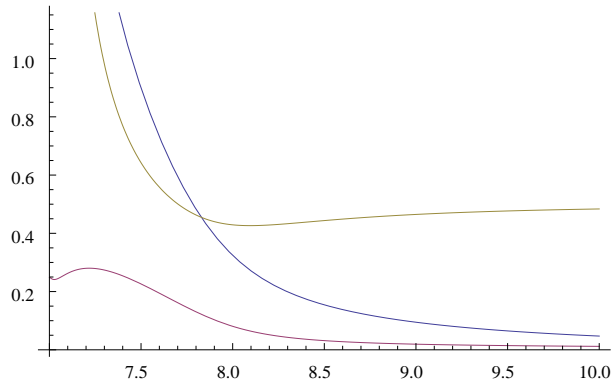}
\caption{Plot of the temperature (curve non-vanishing at infinity), $-g$ (the lowest curve) and $f$ for $f_2=1, T_0=0.5, g_2=-1$. 
The horizontal axis captures the radial coordinate near the divergent core $r_c \approx 6.998$. }
\label{c2nz}
\end{center}
\end{figure}

The most general isolated stationary vortex has both a flux of energy and angular momentum from infinity into a singular sink in its core, corresponding to nonzero values of the constants of integration $c_1$ and $c_2$. We plot a typical profile of the various vortex quantities as a function of the radius in Figure \ref{c2nz}. More exotic profiles are possible: one curious feature of these vortices is that there exist choices of parameters for which the vortex changes its direction of rotation at a finite radius, where it goes through zero rotational velocity. 

{\bf Asymptotic Expansion:} We are interested only in asymptotically static (isolated) vortex solutions. This means that at radial infinity we want the velocity to tend to zero and the temperature to tend to a constant.  It is useful (both for numerics and otherwise) to understand the mutually consistent behavior of the unknown functions at infinity that satisfy the equations of motion. To do this we substitute the expansion
\bea
f&=&
\frac{f_ 2}{r^2}+\frac{f_3}{r^3}+... \\  \label{as1}
g&=&
\frac{g_2}{r^2}+\frac{g_3}{r^3}+... \\
T&=&T_0+\frac{T_1}{r}+\frac{T_2}{r^2}+\frac{T_3}{r^3}+... \label{as2}
\eea
into the ODE form of the equations presented in Appendix B and demand that they be zero at each order. There are no ${\cal O}\big(\frac{1}{r}\big)$ terms for $f$ and $g$ because we want the velocities ($r f$ and $r g$) to die down at infinity. Note that a solution of this form was strictly not necessary\footnote{For example, exponential fall-offs at infinity are not captured by this form.}: but if we do find a solution, then that would be good enough for us (cf. Introduction). We find that the leading constraints\footnote{The subleading constraints can be thought of as equations determining higher order coefficients in terms of lower order ones.} on a consistent solution
\bea
f_1=0, \ \ g_1=0, \ \ f_2=-\frac{9c_2 G}{(4 \pi T_0^2+18 c_1 G )}, \ \ g_2=\frac{9c_1 G}{4 \pi^2 T_0^3}
\eea

A detailed analysis of the phases of the $c_1\neq 0\neq c_2$ case would be interesting, but we will not undertake it here in this preliminary investigation. We demonstrate the existence of a rich phase structure by studying in detail the case where $c_1=0$, corresponding to no energy flow from infinity, in the remainder of this note.

\subsection{$c_1=0$: Large radius expansion}

It is straightforward to perform the asymptotic expansion as in the last subsection for the $c_1=0$ case as well, and fix the fall-offs. But instead we will proceed by direct inspection of the equations at large $r$. (Both methods lead to identical results of course.) First, inspired by the incompressible vortex presented earlier, we look for a rotational velocity $rf$ that decreases as $1/r$ while $g$ decreases faster at large $r$.  This means that the angular velocity $f$ decreases as $1/r^2$
\beq
f=\frac{f_2}{r^2}
\eeq
where $f_2$ is a constant. The vorticity of a vortex is the integral of the curl of the velocity.  In an ideal incompressible (2+1)-dimensional vortex, this is a delta function at the origin, and one may use Stokes' theorem to evaluate it as the integral of the rotational velocity along a fixed radius circle.  We will adopt this terminology, and refer to the vorticity of a vortex as this integral, evaluated far enough from the core that the nonrelativistic approximation holds.  The rotational velocity is then equal to $f_2/r$ and so the vorticity is
\beq
\omega=2\pi f_2. \label{omega}
\eeq

Now let us attempt to determine the leading order $r$-dependence of the energy conservation condition (\ref{alpha}).  $P$ and $\eta$ asymptote to constants due to asymptotic ideality.  $\gamma$ tends to 1, as the velocity goes to zero.  Thus the first term on the left hand side has the same $r$ dependence as $g$, while the second term has the same $r$ dependence as $A$, which is expressed in terms of $f$ and $g$ in Eq.~(\ref{AA}).  The dominant term at large r is
\beq
A\sim rff\p\sim -\frac{2f_2^2}{r^4} \label{aasi}
\eeq
which is of order $1/r^4$.  On the other hand, the right side of Eq.~(\ref{alpha}) is $c_1/r^2$, which is of lower order in $r$ than either $A$ (which is of order $1/r^4$) or $g$ (which falls off faster than $1/r^2$).  In order for (\ref{alpha}) to be satisfied, the order $1/r^2$ term on the right hand side must vanish, and so we learn that
\beq
c_1=0
\eeq
as desired.  

Now that $c_1=0$, the two terms in Eq.~(\ref{alpha}) must be equal
\beq
3\gamma g P=\frac{\eta}{\gamma} A. \label{a2}
\eeq
As $P$, $\eta$ and $\gamma$ all tend to constants at large $r$, $g$ must have the same scaling as $A$
\beq
g\sim \frac{g_2}{r^4}.
\eeq
The constant $g_2$ is nonzero and in fact negative, as $\eta$ and $\gamma$ are nonzero and positive, and if the vorticity $\omega$ is nonzero so is $f_2$ by (\ref{omega}) and therefore $A$ is strictly negative by Eq.~(\ref{aasi}).  Thus we have found that 
the Landau frame velocity will always have an inward component, similarly to the ordinary velocity in a stretching Burgers vortex.

From (\ref{a2}) and (\ref{aasi}) we see that this inward velocity is proportional to the viscosity and the rotational velocity squared and inversely proportional to the energy density.  This is reasonable for the following reason.  In the presence of shear viscosity, there is a net flow of angular momentum/energy from regions of smaller to greater tangential velocity: viscosity couples to velocity gradients.  As the velocity decreases with radius, this means that energy-momentum flows from smaller to larger radii.  
In a stationary configuration, this outward energy flow must be canceled.  It is canceled by an inward velocity.  The inward energy flow is equal to the inward velocity times the energy density $\rho=2P$.  
The content of (\ref{a2}) is precisely the balancing between these two terms.

We may furthermore use (\ref{a2}) to find the ratio of the pressure and the viscosity
\beq
\frac{P}{\eta}=\frac{A}{3\gamma^2g}.
\eeq
In Eq.~(\ref{P1}) we have already seen that this ratio may be expressed in terms of the temperature
\beq
\frac{P}{\eta}=\frac{4\pi T}{3}=\frac{A}{3\gamma^2g}
\eeq
and so we may find the temperature in terms of the velocity
\beq
T=\frac{A}{4\pi\gamma^2g}. \label{temp}
\eeq
This expression for $T$ is exact, it is not the asymptotic value of $T$ at large $r$.  Such an exact result is possible because, once we have established that $c_1$ vanishes at large $r$, it vanishes everywhere.  We may use it together with (\ref{aasi}) to determine the asymptotic value $T_0$ of the temperature in terms of the asymptotic velocities
\beq
T_0=\frac{-f_2^2}{2\pi g_2}. \label{T0}
\eeq
This is positive as $g_2$ is negative.

Note that the temperature is independent of Newton's constant $G$, it is written purely as a function of the velocities in the problem.  However, as we used $\eta$ and $P$ to arrive here, in conformal fluids without gravity duals this expression for $T$ will generally be corrected by a multiplicative factor.

Now that $T$ has been expressed as a function of the velocities, it is easy to use Eqs.~(\ref{P1}, \ref{P2}) to express the other thermodynamic variables as functions of the velocities
\beq
\eta=\frac{A^2}{144\pi G\gamma^4g^2}\hsp
P=\frac{A^3}{432\pi G\gamma^6 g^3}.  \label{p2}
\eeq
One then finds the asymptotic values of $\eta$ and $P$
\beq
\eta_0=\frac{f_2^4}{36\pi G g_2^2}\hsp
P_0=\frac{-f_2^6}{54\pi G g_2^3}
\eeq
where again $P_0$ is positive because $g_2$ is negative.

Plugging the expressions (\ref{p2}) back into the equations of motion (\ref{alpha},\ref{calc},\ref{cald},\ref{cale}) one finds a system of differential equations that depends only the velocities, with no thermodynamic quantities.  Moreover, by choosing $\eta$ and $P$ as in Eq.~(\ref{p2}) we have satisfied the conservation of energy condition (\ref{alpha}).  Eq.~(\ref{calc}) is the definition of $\mathcal{C}$ as a function of $f$, $g$ and their first derivatives, and thus may also be eliminated.  Eq.~(\ref{cald}) depends on the first derivative of $\mathcal{C}$ and so is a second order ordinary differential equation in the velocities, while Eq.~(\ref{cale}) is a first order differential equation.  Thus we are left with one ordinary first order differential equation and one ordinary second order differential equation for two functions $f(r)$ and $g(r)$.  One then expects that solutions will be parametrized by $c_2$ and 3 constants of integration.  One of these is given by the vorticity $\omega=2\pi f_2$ and one by the asymptotic temperature $T_0$. The various phases of the solutions only depend on the dimensionless combination $\omega T_0$.  
The fact that two of the constants of integration do not lead to parameters in our space of solutions is not surprising because typically we expect also growing modes to be present which are outside of our asymptotic fall-off ansatz. In our case in fact it seems also likely that the two missing modes might in fact be a pair of complex (conjugate) solutions for at least some regimes (of $r$ and parameters) due to some suitably defined discriminant vanishing. We have found such complex solutions analytically for various special cases\footnote{One of which we report in Subsec.~\ref{nullsec}.}: they involve functions of the form $\sqrt{C^2-r^2}$ which are not well-defined for $r \ge C$ for some integration constant $C$. We have not attempted a general analysis of such solutions because our interest is primarily in asymptotically ideal solutions.

The constant $c_2$ is easily evaluated in terms of the boundary conditions $f_2$ and $T_0$ and the conservation of angular momentum (\ref{cale}).  First one needs to find the asymptotic behavior of $E$ from Eq.~(\ref{EE}).  It is dominated by the $f\p/r$ term
\beq
E\sim \frac{-2f\p}{r}\sim\frac{4f_2}{r^4}.
\eeq
Substituting this into (\ref{cale}) one finds
\beq
c_2=-r^4(6fgP+\eta E)\sim -r^4\eta E\sim -4\eta_0 f_2=-\frac{f_2^5}{9\pi G g_2^2}.
\eeq

\subsection{$c_1=0$: Vortices with bounded velocities} \label{bergersec}

As we report in Sec.~\ref{numsec}, numerically we have found two different kinds of solutions in the $c1=0\neq c2$ case, characterized by different behaviors in the core of the vortex.  One kind of solution, which occurs for $\omega T_0\lesssim .1$, is characterized by a bounded 3-velocity $u$.  In this subsection we will analytically determine the leading order behavior of this solution at small values of $r$.

The inward velocity $rg$ tends to a constant value $u_{\rm{rad}}$ as $r\rightarrow 0$ while the rotational velocity $rf$ tends to zero as a power of $r$
\beq
f\sim f_\delta r^\delta\hsp
g\sim \frac{u_{\rm{rad}}}{r}\hsp
\delta>-1.
\eeq
Recall that our differential equations for the velocity depend only on $f$ and $g$, and so this is enough information to determine the velocities and also the temperature by Eq.~(\ref{temp}).  The pressure and viscosity may also be determined in terms of the dual Newton's constant $G$.

One may now calculate the small $r$ behavior of all of the functions that we have introduced.  As the rotational velocity tends to zero, its contribution to $\gamma$ also tends to zero leaving
\beq
\gamma\sim\sqrt{1+u_{\rm{rad}}^2}
\eeq
and so not only does the relativistic inward velocity tend to a constant, but so does the true inward velocity $u/\gamma$.  Similarly the true rotational velocity tends to zero.

The leading order limiting behaviors of the other functions follow similarly
\bea
&&A\sim -\frac{(1+u_{\rm{rad}}^2)u_{\rm{rad}}^2}{r^2}\hsp
C\sim  -\frac{(1+u_{\rm{rad}}^2)u_{\rm{rad}}}{r^3}\hsp
D\sim  \frac{u_{\rm{rad}}}{r^3}\hsp
E\sim -2(u_{\rm{rad}}^2+\delta+\delta u_{\rm{rad}}^2)f_\delta r^{\delta-2}\nonumber\\
&&\mathcal{C}\sim \frac{u_{\rm{rad}}^3}{216\pi G r^5}\hsp
\mathcal{D}\sim -\frac{u_{\rm{rad}}^3}{108\pi G r^5}\hsp
\mathcal{E}\sim (-4f_\delta u_{\rm{rad}}^4-2(\delta +\delta u_{\rm{rad}}^2)f_\delta u_{\rm{rad}}^2)r^{\delta-4}. \label{sviluppi}
\eea
Notice that Eq.~(\ref{cald}) is satisfied, thus we need only check that angular momentum is conserved (\ref{cale}).  This equation implies that the leading order term in $\mathcal{E}$ is of order $1/r^4$ with a coefficient which is nonvanishing for a vortex with $\omega\neq 0$, in other words, for any vortex.  Here we have found a leading order term of order $r^{\delta-4}$.  Therefore $\delta$ may not be greater than 0, because there would be no order $1/r^4$ term.

Thus we are left with the condition
\beq
-1<\delta<0. \label{rango}
\eeq
Numerically we find that vortex solutions exist with all values of $\delta$ in this range.  Now the leading term in $\mathcal{E}$ must be order $1/r^4$, whereas the term in (\ref{sviluppi}) is of higher order.  Therefore this higher order coefficient must be zero.  Dividing through by $-2f_\delta u_{\rm{rad}}^2$ this implies
\beq
2u_{\rm{rad}}^2+\delta u_{\rm{rad}}^2+\delta=0
\eeq
and so
\beq
u_{\rm{rad}}^2=-\frac{\delta}{2+\delta}. \label{gd}
\eeq
In particular the range of delta (\ref{rango}) determines the range of $u_{\rm{rad}}$ and so $\gamma$
\beq
-1 < u_{\rm{rad}} < 0
\hsp
0<\gamma<\sqrt{2}
\eeq
where we have chosen a negative sign for $g_0$ in order to obtain a positive temperature.
Thus the maximum true velocity $u/\gamma$, which occurs for a vortex at the critical vorticity
\beq
\omega_c\sim \frac{0.1}{T_0} \label{oc}
\eeq
is only $1/\sqrt{2}$ times the speed of light.  On the other hand, the velocity at the core of the vortex may be relatively small if $\omega$ is taken small enough and need not even be relativistic, although it is discontinuous at the origin.

Although the velocity does not diverge, the thermodynamic quantities do diverge.  Eq.~(\ref{temp}) and Eq.~(\ref{p2}) imply that to leading order they diverge as
\beq
T\sim\frac{-u_{\rm{rad}}}{4\pi r}\hsp
\eta\sim\frac{u_{\rm{rad}}^2}{144\pi Gr^2}\hsp
P\sim\frac{-u_{\rm{rad}}^3}{432\pi Gr^3}.
\eeq
Again these quantities are positive because $u_{\rm{rad}}$ is negative.  These divergences are features of our stationary ansatz, and so are not expected to persist in vortices that have formed over a finite period of time.  

We have determined the leading order behavior of the solution at the origin in terms of the unknown variables $u_{\rm{rad}}$ and $f_\delta$, where $f_\delta$ is increasingly irrelevant as it is the coefficient of the rotational velocity which tends to zero at the origin.  Being dimensionless, $u_{\rm{rad}}$ may only depend on the dimensionless combination of constants of integration $\omega T_0$, we will write this fact as
\beq
u_{\rm{rad}}=u_{\rm{rad}}(\omega T_0).
\eeq
We are unable to determine this function analytically, and so will find it numerically in Fig.~\ref{urad} of Sec.~\ref{numsec}.

\subsection{$c_1=0$:   Vortices with a finite radius divergence} \label{rfin}

For a given value of the asymptotic temperature $T_0$, if the vorticity $\omega$ is greater than the critical vorticity $\omega_c$ in (\ref{oc}), at a finite radius $r_c$ the rotational velocity $r_cu$ is infinite, and so the true velocity is the speed of light.

Numerically we have found that the behavior near $r_c$ is universal.  Rather than demonstrating that no other behavior is allowed, we will merely show that the behavior that we find satisfies the equations of motion.  When $r$ is slightly greater than $r_c$ we find
\beq
f\sim \frac{2}{\sqrt{r_c}\sqrt{r-r_c}}\hsp
g\sim -\frac{\sqrt{3}}{r_c}. \label{muro}
\eeq
In particular, the angular velocity diverges.  In this case not only the powers of the leading terms in the velocity but even the coefficients themselves are entirely determined by $r_c$.  $r_c$ has dimensions of length, like the vorticity $\omega$ and so $r_c/\omega$ must be entirely determined by the only dimensionless combination of the constants of integration $\omega T_0$.  Therefore
\beq
r_c=\omega h(\omega T_0) \label{h}
\eeq
for some function $h$, which is plotted in Fig.~\ref{rc} of Sec.~\ref{numsec}.

Now that we have formulas for $f$ and $g$, we know the limiting behaviors of the velocities and so, as in the previous subsection, may determine the limiting behaviors of the other functions.  This time $f$ dominates over $g$, and so $\gamma$ is dominated by the large rotational velocity
\beq
\gamma\sim r_cf\sim \frac{2\sqrt{r_c}}{\sqrt{r-r_c}}.
\eeq
In particular $\gamma$ diverges and so the true inward velocity $rg/\gamma$ tends to zero.   The limiting true rotational velocity $rf/\gamma$ tends to the speed of light.

The other functions are easily found to be
\bea
&&A\sim-\frac{24}{(r-r_c)^2}\hsp
C\sim\frac{8\sqrt{3}}{r_c^2(r-r_c)}\hsp
D\sim\frac{4\sqrt{3}}{r_c(r-r_c)^2}\hsp
E\sim\frac{24}{r_c^{3/2}(r-r_c)^{3/2}}\nonumber\\
&&\mathcal{C}\sim-\frac{1}{3\sqrt{3}\pi G r_c^2(r-r_c)^3}\hsp\hspace{-.3cm}
\mathcal{D}\sim\frac{1}{\sqrt{3}\pi G r_c^2(r-r_c)^3}\hsp\hspace{-.3cm}
\mathcal{E}\sim\frac{2-2}{\pi G r_c^{3/2}(r-r_c)^{9/2}}. \label{muroas}
\eea
Here the vanishing of the $O((r-r_c)^{-9/2})$ term of $\mathcal{E}$ is expressed as above to show that, with the particular values of the constants in (\ref{muro}), this leading term, which would be inconsistent with the equation of motion Eq.~(\ref{cale}), vanishes.  In principle, to verify that (\ref{cale}) is satisfied with a nonvanishing value of $c_2$, one should also calculate the order $1/r^4$ term, but we have simply verified numerically that this works.  One can see from (\ref{muroas}) that the other equation of motion (\ref{cald}) is satisfied.

The thermodynamic quantities are also divergent at $r_c$.  The temperature diverges as
\beq
T\sim \frac{\sqrt{3}}{2\pi(r-r_c)}
\eeq
while the viscosity and pressure diverge as
\beq
\eta\sim\frac{1}{12\pi G (r-r_c)^2}\hsp
P\sim\frac{1}{6\sqrt{3}\pi G (r-r_c)^3}.
\eeq
The divergence of thermodynamical quantities in the infinitely boosted Landau rest frame of the liquid may not be particularly physically relevant.  However the energy $T^{tt}$ in the rest frame of the asymptotic fluid diverges even more strongly
\beq
\sigma^{tt}\sim\frac{4\sqrt{3}r_c}{(r-r_c)^2}\hsp
T^{tt}\sim\frac{r_c}{\sqrt{3}\pi G (r-r_c)^4.}
\eeq
In fact there is nearly a cancellation leading to a lesser divergence, the viscosity term in the energy is minus one half of the pressure term. This semi-cancellation has some of the flavor of a Virial theorem.

\subsection{$c_1=0=c_2$: No asymptotically static vortices} \label{nullsec}

Finally we turn to the case $c_1=c_2=0$, in which there still is no energy flux but now also no angular momentum flux. It is easy to see from the asymptotic behavior of the first order form of the equations in Appendix B that no asymptotically static solutions can exist (with non-zero temperature at infinity) in this case. But there is a slight subtlety to this statement. This is because in going from the original equations to the first order form, we have divided in places by $g$ and so we are implicitly assuming that $g \neq 0$. This would have been a mere technicality had it not been for the fact that in the viscosity $\rightarrow 0$ limit we did indeed find vortex solutions with $g=0$ in (\ref{nonvisc}). 
This suggests that we look for an exact solution with $g=0$. In fact, if one works with the original form of the equations of motion in the $c_1=c_2=0$ case and sets $g=0$ and looks for solutions, we can find the full analytic solution:
\bea
\frac{T'}{T}\equiv f^2 r, \ \ {\rm with} \ \ f \sim \frac{1}{\sqrt{C^2-r^2}}
\eea
for some constant $C$. It is easy to see that this solution is not asymptotically static.


\section{Numerical Results} \label{numsec}

\begin{figure}
\begin{center}
\includegraphics[scale=1.08]{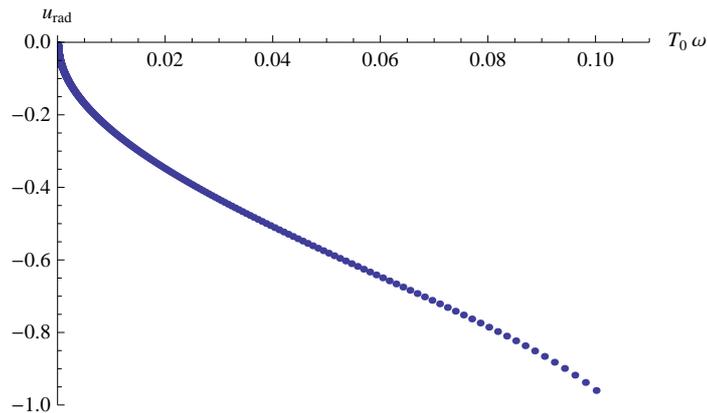}
\caption{When $\omega<\omega_c$, the velocity is everywhere bounded.  The rotational velocity tends to zero at the origin, and the angular velocity diverges as $r^\delta$.  The leading order behavior may be entirely determined by $u_{\rm{rad}}$, which in turn is a function of the dimensionless combination $\omega T_0$.  In this figure we numerically determine this function.  Notice that at the critical value of $.1$ the coefficient $u_{\rm{rad}}$ goes to $-1$, which we have seen is the maximum velocity allowed for a solution with this scaling.  Beyond this critical value the rotational velocity diverges at a finite radius $r_c$, and the solutions are in the other phase.}
\label{urad}
\end{center}
\end{figure}

\begin{figure}
\begin{center}
\includegraphics[scale=1.08]{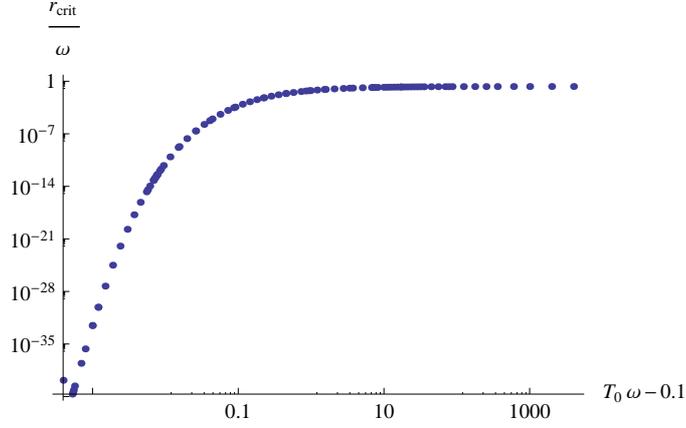}
\caption{When $\omega>\omega_c$, the velocity reaches the speed of light at a finite critical radius, $r_{\rm{crit}}$.  The dimensionless combination $r_{\rm{crit}}/\omega$ is a universal function of the dimensionless combination $\omega T_0$.  This function is plotted above.  As expected, the critical radius tends to zero as $\omega$ approaches $\omega_c$, in accordance with the fact that the velocity is bounded when $\omega<\omega_c$.  On the other hand at a fixed temperature in the high vorticity limit, the critical radius asymptotes to the vorticity.  In particular for all solutions the critical radius is bounded by the vorticity.}
\label{rc}
\end{center}
\end{figure}

\begin{figure}
\begin{center}
\includegraphics[scale=.88]{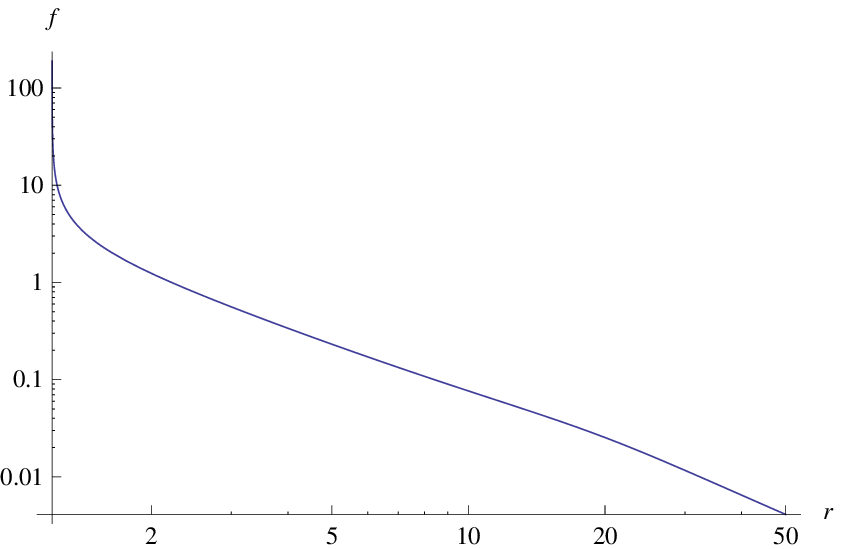}
\includegraphics[scale=.88]{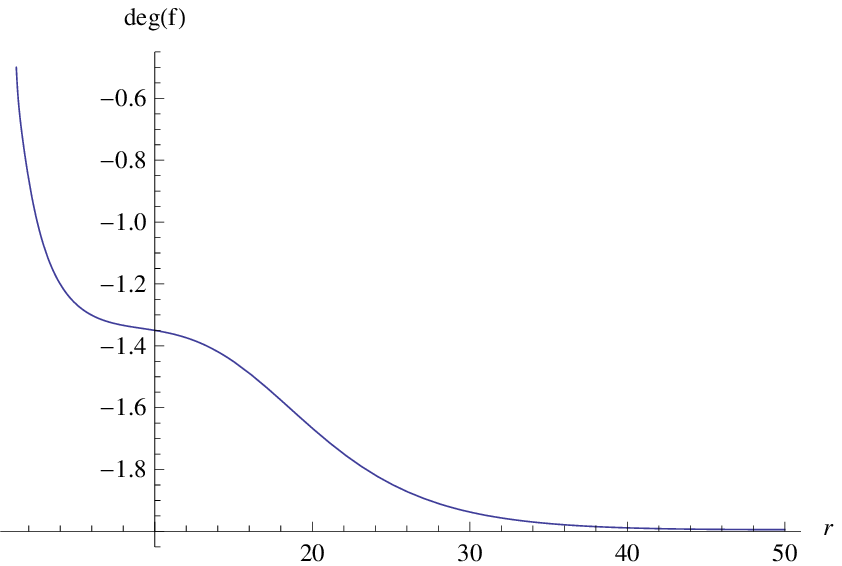}
\caption{This is the profile of the angular velocity $f$ versus radius for the vortex characterized by the boundary conditions $f_2=10$ and $\omega T_0=0.5$.  The rotational velocity reaches the speed of light at $r_c\sim 1.208$.  At large radii the angular velocity scales as $1/r^2$, and near $r_c$ it scales as $(r-r_c)^{-1/2}$.  On the right we display the degree of $f$, defined as the exponent of $(r-r_c)$ in its leading term, or more precisely as $(r-r_c)f\p/f$. }
\label{murf}
\end{center}
\end{figure}

\begin{figure}
\begin{center}
\includegraphics[scale=.88]{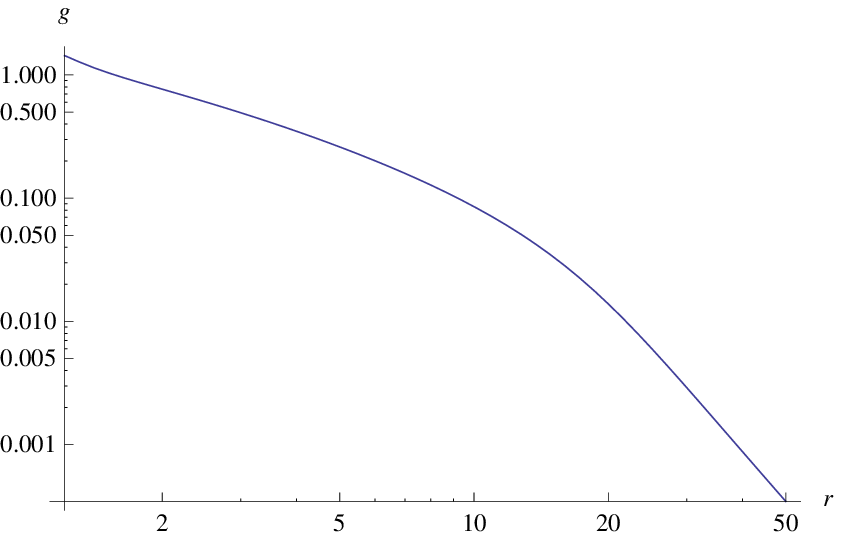}
\includegraphics[scale=.88]{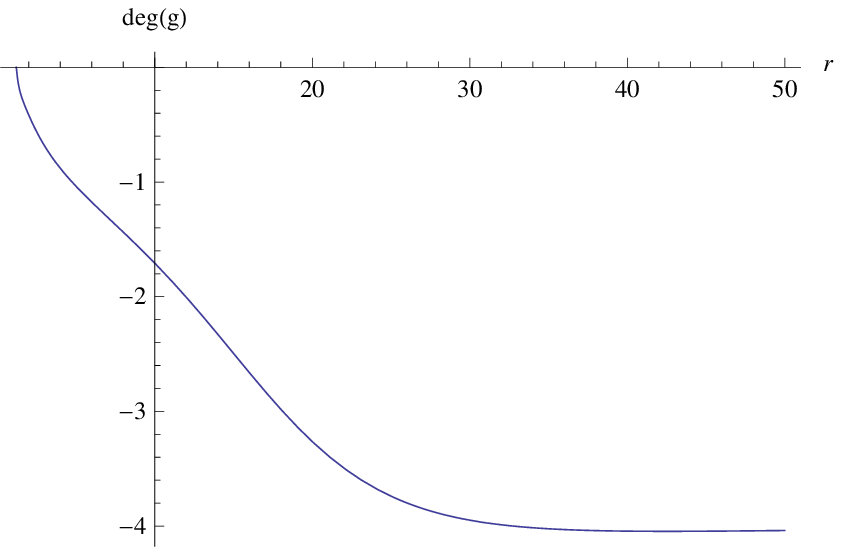}
\caption{This is the profile of the fractional contraction velocity $-g$, which is the true inward radial velocity times $\gamma$ divided by the radius, versus the radial coordinate.  The vortex is characterized by the boundary conditions $f_2=10$ and $\omega T_0=0.5$.  While the rotational velocity reaches the speed of light at $r_c\sim 1.208$, the radial velocity tends to a constant.  At large radii the fractional expansion velocity scales as $1/r^4$, corresponding to a radial velocity that scales as $1/r^3$.  On the right we display the degree of $g$, defined as the exponent of $(r-r_c)$ in its leading term, or more precisely as $(r-r_c)g\p/g$.} 
\label{murg}
\end{center}
\end{figure}

\begin{figure}
\begin{center}
\includegraphics[scale=.88]{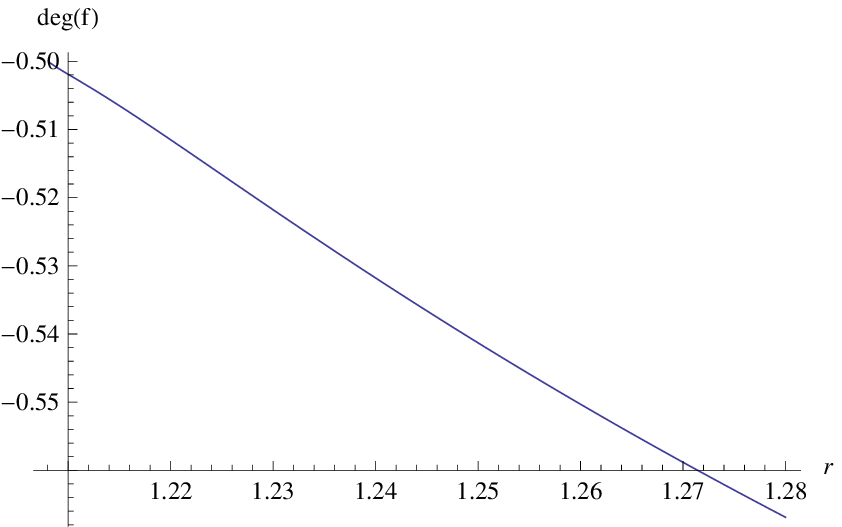}
\includegraphics[scale=.88]{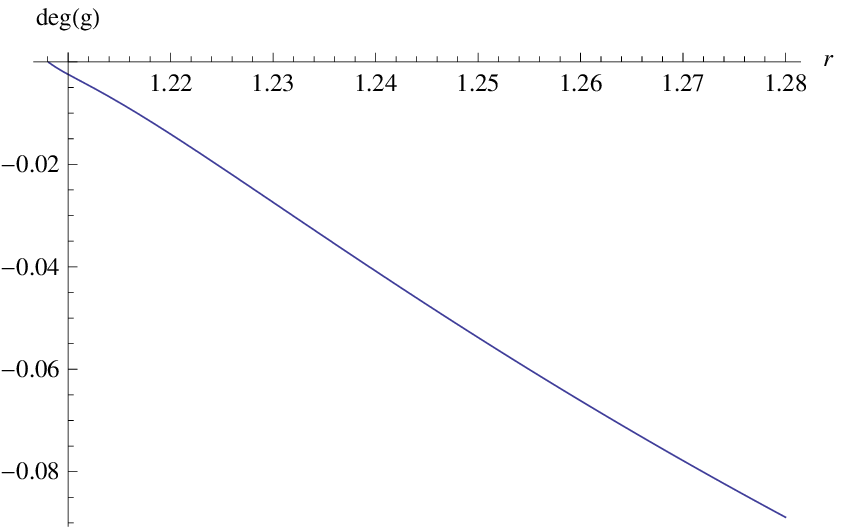}
\caption{This is the profile of the degrees of the angular and radial velocity profile functions $f$ and $g$ near the critical radius $r_c\sim 1.208$.  These degrees are defined as $(r-r_c)f\p/f$ and $(r-r_c)g\p/g$.  The vortex is characterized by the boundary conditions $f_2=10$ and $\omega T_0=0.5$.  One can see that the exponents of the $(r-r_c)$ scaling tend to the well-defined limits $-0.5$ and $0$ found in the asymptotic analysis.}
\label{murvic}
\end{center}
\end{figure}

\begin{figure}
\begin{center}
\includegraphics[width=8cm]{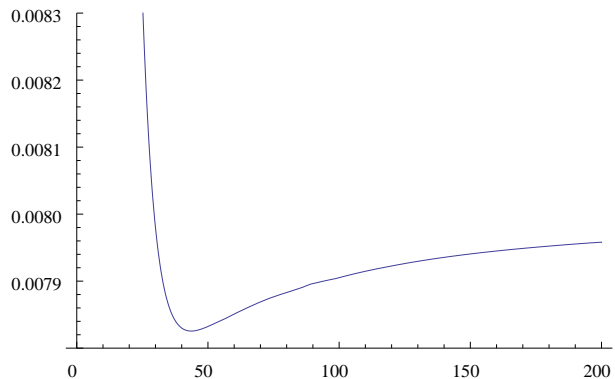}
\caption{This is the radial temperature profile of the vortex with $\omega T_0=0.5$ and $f_2=10$.  The temperature diverges at $r_c$, but at large $r$ tends to a constant, with corrections of order $1/r^2$. 
}
\end{center}
\end{figure}

Now we turn to a numerical investigation of the $c_1=0$ vortices discussed above.  Recall that in Subsecs.~\ref{bergersec} and \ref{rfin} we found two phases of solutions.  The first phase is characterized by a finite inward velocity $u_{\rm{rad}}$ at the origin, while the rotational velocity tends to zero as $r^{1+\delta}$.  By dimensional analysis  $u_{\rm{rad}}$ may be written as a function of the dimensionless combination $\omega T_0$.  In Fig.~\ref{urad} we numerically find this function and we find that, as we have claimed, this phase is manifested when $\omega T_0\lesssim 0.1$.  In particular, using Eq.~(\ref{gd}), we see that as one approaches the critical vorticity $\omega_c=0.1/T_0$, $\delta$ tends linearly to $-1$.  This is to be expected, it implies that at the critical vorticity the rotational velocity no longer tends to zero.  In fact, one reaches the limiting case $r_c=0$ of the second phase, in which the relativistic rotational velocity tends to infinity at $r=r_c$.  The critical radius of this phase, in turn, is plotted in Fig.~\ref{rc}.

We also display the profiles of some typical vortices in these two phases.  First, we display graphical results for the numerical solution corresponding to $f_2=10$ and $\omega T_0=0.5$.  As $\omega T_0$ is five times the critical value, the velocity will reach the speed of light at a finite radius $r_c$.  In this case $r_c\sim 1.208$.  The angular velocity $f$ profile is displayed in Fig.~\ref{murf}.  We see that, in agreement with the asymptotic analysis in Sec.~\ref{assec}, at large $r$, $f$ tends to $f_2/r^2$, corresponding to a rotational velocity of $f_2/r$.  As $r$ approaches $r_c$, $f$ diverges as $1/\sqrt{r-r_c}$, corresponding to an angular velocity that approaches the speed of light.  The radial velocity is seen in Fig.~\ref{murg}.  Again, in agreement with the asymptotic analysis, the fractional radial velocity scales as $1/r^4$ at large $r$, corresponding to a radial velocity that scales as $1/r^3$.  At large values of the radius this solution indeed becomes the usual nonrelativistic solution.  Near the critical radius $r_c$, $g$ tends to a constant.  This corresponds to a constant inward relativistic radial velocity.  However, as $\gamma$ diverges, the true radial velocity tends to zero.  In Fig.~\ref{murvic} we zoom in on the region near $r_c$, seeing that the scaling exponents of $f$ and $g$ indeed tend to $-0.5$ and $0$ as claimed in Subsec.~\ref{rfin}.


The only dimensionful quantities that are needed to characterize the solutions are $f_2$ and $r_c$, therefore if $f_2$ is varied with $\omega T_0$ fixed, all of these results must remain the same except that $r_c$ will change such that $f_2/r_c$ is constant.  Indeed this is the content of Eq.~(\ref{h}).  We have numerically verified that this is indeed the case.

\begin{figure}
\begin{center}
\includegraphics[scale=.88]{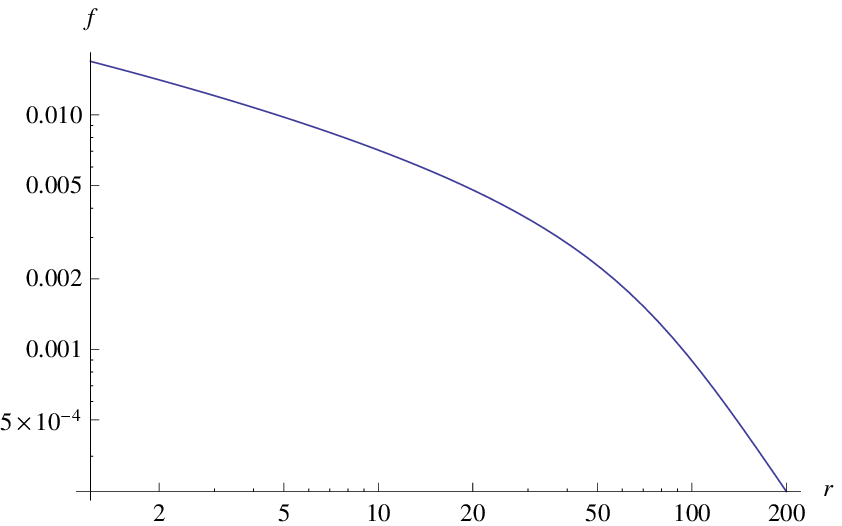}
\includegraphics[scale=.88]{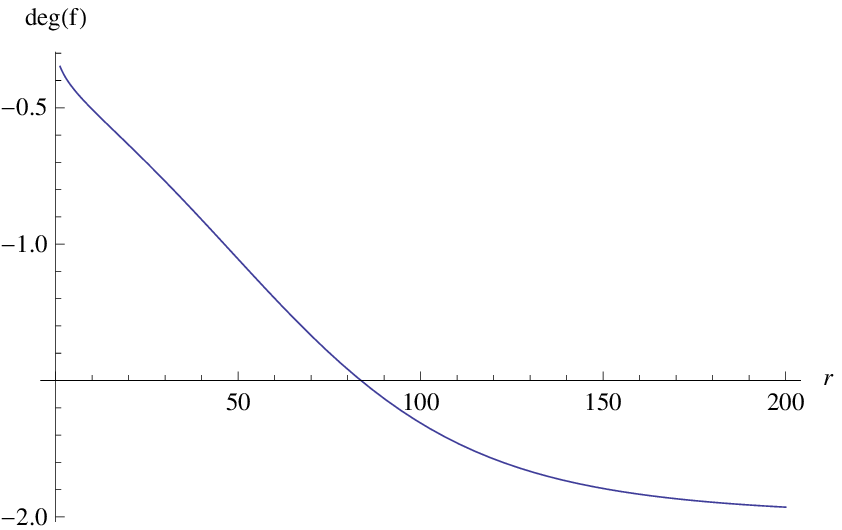}
\caption{This is the profile of angular velocity $f$ versus radius for the vortex characterized by the boundary conditions $f_2=10$ and $\omega T_0=0.02$, which is one fifth of the critical value.  The rotational velocity vanishes at the origin as $r^{0.78}$, and as always  at large radii the angular velocity scales as $1/r^2$.}
\label{bergf}
\end{center}
\end{figure}

\begin{figure}
\begin{center}
\includegraphics[scale=.88]{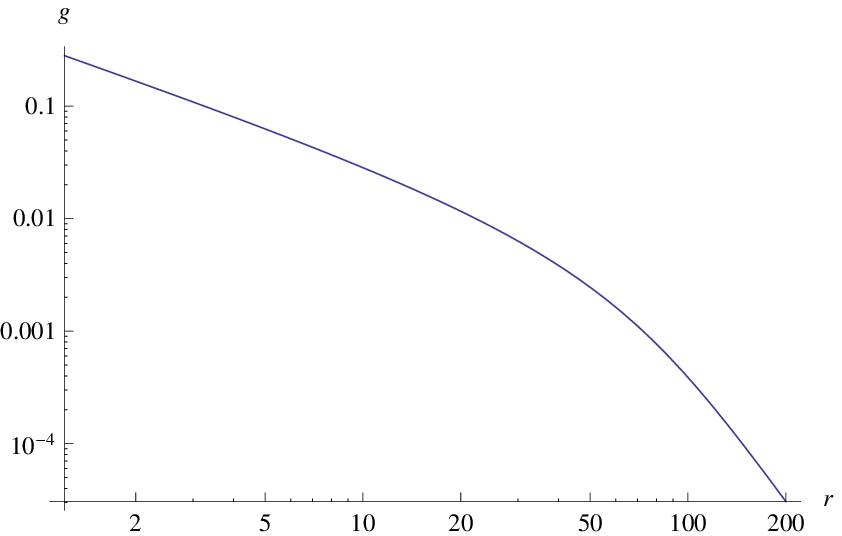}
\includegraphics[scale=.88]{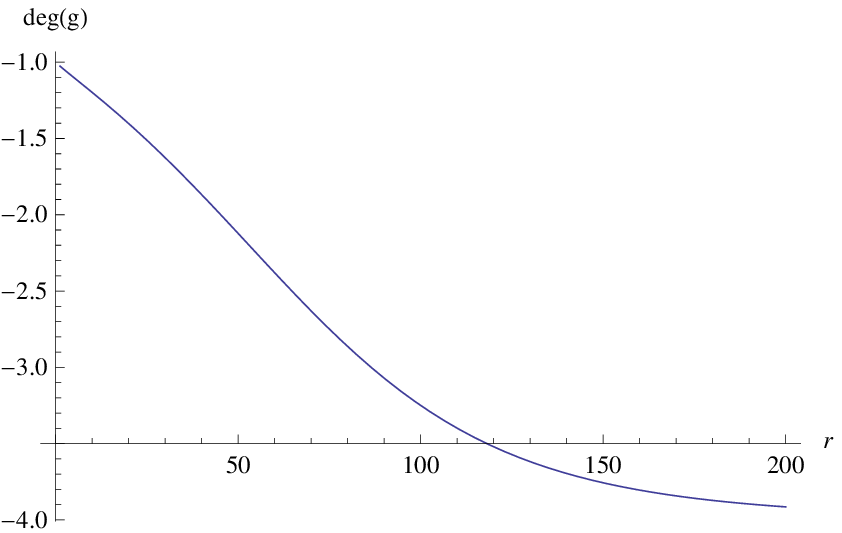}
\caption{This is the profile of the fractional contraction velocity $-g$ versus the radius for the vortex characterized by the boundary conditions $f_2=10$ and $\omega T_0=0.02$.  At the origin the inward velocity $-rg$ tends to a constant value of $0.3485$, while asymptotically it falls as $1/r^3$ as always.}
\label{bergg}
\end{center}
\end{figure}

\begin{figure}
\begin{center}
\includegraphics[width=8cm]{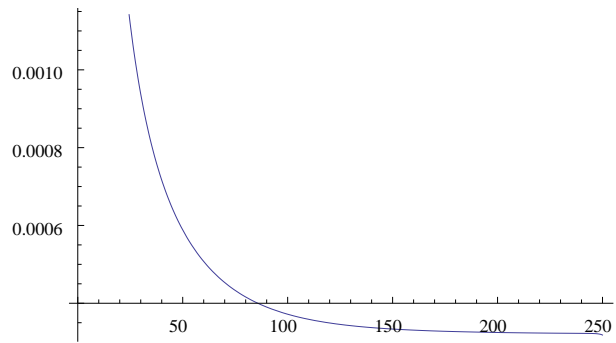}
\caption{This is the radial temperature profile of the vortex with $\omega T_0=0.02$ and $f_2=10$.  The temperature diverges at the origin, but at large $r$ tends to a constant.}
\label{tempfig}
\end{center}
\end{figure}

Finally we display a case in which the dimensionful parameter $f_2$ is the same, but the dimensionless combination of asymptotic parameters is $\omega T_0=0.02$, which is one fifth of the critical value.  Therefore the velocity remains everywhere bounded.  In Fig.~\ref{bergf} we plot the angular velocity $f$ versus radius, seeing again that at large $r$ it scales as $1/r^2$, but that at at small $r$ it scales as $r^{-0.22}$, indicating that the rotational velocity $rf$ falls to zero as $r^{0.78}$ near the origin.  The radial fractional velocity $g$ is displayed in Fig.~\ref{bergg}.  Again asymptotically it shrinks as $1/r^4$.  Now near the origin it diverges as $1/r$, implying that the inward velocity $rg$ tends to a constant, which is in this case $0.349$.  Thus $\delta=-0.22$ and the small $r$ coefficient $u_{\rm{rad}}=0.349$, satisfying Eq.~(\ref{gd}).  All of these results, being dimensionless, must not change if $f_2$ is varied with the dimensionless combination $\omega T_0$ fixed.  We have checked numerically that this is indeed the case.  In Fig.~\ref{tempfig} we see the radial temperature profile in vortex with $\omega T_0\sim .02$.  Consistently with the asymptotic analysis we see that the temperature tends to a constant at large $r$, with $1/r^2$ corrections while it blows up at small $r$.

\section{Discussion}

Finding a vortex solution is clearly the most preliminary of steps in an approach to understanding the dynamics of a fluid and in particular, turbulent flow. We have investigated 2+1 dimensional vortices as a warm-up example: the 
intersting case 
of 3+1 dimensions 
should also be tractable. It would be interesting to see if one can find a solution that generalizes Burgers vortex to 3+1 conformal fluids. This might also be important in understanding real-world systems such as layers of neutron stars which are believed to be well-approximated by very low-viscosity conformal fluids. 

One of the questions that we have not investigated is that of the  stability of the vortices.  The fact that numerics is robust suggests that at the first order derivative expansion for the stress tensor that we are working with, the solutions are likely to be stable. At the core of the vortex where the quantities blow up, a regular solution might be possible by going beyond the first order terms. But as we have mentioned, stationary vortices serve as ingredients in a turbulent, time-dependent flow where singularities are avoided. 
In incompressible fluids,  the full flow is expected to be understood purely using the second order  (in the EOM, i.e., Navier-Stokes)  terms for regular initial data. It will be interesting to determine for which regimes this approximation holds for conformal flows.

An interesting aspect of the stability question is that due to the fluid-gravity correspondence, it could translate to questions of stability in black hole spacetimes. In fact, by taking a scaling limit of a conformal fluid flow, in \cite{MinNR}, a map between the turbulent instabilities of Navier-Stokes at high Reynolds numbers and gravitational solutions was made. But there is a tautological nature to the fluid-gravity translation: for any flow satisfying certain general conditions one can write down a gravity solution order by order in derivatives. It is less clear at the moment how to put this to work and use one side of the duality to understand the long-standing questions in the other. Some related discussion on stability, higher derivative terms and regularity of flows can be found in the Introduction. Shock waves and boundary effects have also been discussed in this context in \cite{Krucz, Ricco}.

\subsection{Non-Relativistic Limit}

Asymptotically, the fluids we consider are dual to static black branes. The fact that the temperature is tending to a constant means that the (relativistic) density is also going to a constant.  In fact it is easy to check that in the asymptotic region, our vortex solutions $f$ and $g$ also satisfy the incompressible Navier-Stokes equations up to corrections that die down at infinity. Using the incompressible fluid equations that we wrote down in section 2, for the ansatz (\ref{inans}) the incompressibility and Navier-Stokes equations take the form
\bea
{\rm Incompressibility} \  \equiv 2g +r g' =0 \hspace{1in}\\
{\rm Navier-Stokes} \equiv  r g (2f +r f')'- \frac{\nu}{r}\frac{d}{dr}\Big(r(2f+r f')\Big)=0.
\eea
It is easily checked that for the {\em relativistic} vortex solutions we found for the case $c_1\neq 0 \neq c_2$, in the asymptotic region
\bea
{\rm Incompressibility}\sim{\cal O}\Big(\frac{1}{r^2}\Big), \ \
{\rm Navier-Stokes} \sim {\cal O}\Big(\frac{1}{r^4}\Big),
\eea
while for the case when $c_1=0\neq c_2$, both equations are order $1/r^4$. The velocities in both cases are of the order $1/r$.
Thus, our vortex solutions can be thought of as the UV (near-core) corrections to the incompressible vortex solutions. It will be interesting to see whether this is related to a certain scaling (``non-relativistic'') limit that was taken in \cite{Oz1} to reproduce the incompressible Navier-Stokes equation from conformal fluid dynamics. 

It is important to keep in mind that a constant density regime of the flow does not automatically mean that the relativistic nature of the underlying fluid can be ignored \cite{Raamsdonk}. The reason for this is that in non-relativistic fluids, the pressure is always negligible compared to the energy density while this is not true for relativistic fluids. Our claim above is that the fluid velocities satisfy the incompressible equations of motion (up to some higher corrections in $1/r$) as can be checked by direct computation.

It should be noted that the question of what precisely constitutes a ``non-relativistic'' limit is somewhat ambiguous for relativistic fluids. In  conventional fluid dynamics \cite{Landau}, the fluid particle is a black box with an equation of state and (local) thermodynamics, but with no microscopic structure. In particular, it is entirely consistent to view the thermodynamic variables like pressure, energy density and temperature as god-given quantities with no microscopic origin. This means that when one takes the slow-motion, i.e. non-relativistic, limit for fluid velocity ($\equiv$ velocity of a fluid particle), one has no canonical way of taking the non-relativistic limit of thermodynamical variables like pressure an temperature. To emphasize this, in an appendix we demonstrate a non-relativistic limit for barotropic fluids where the equation of state before and after the limit stays unchanged. This should be contrasted with the limit defined in \cite{Oz1} (for the special case of conformal barotropic fluids), which changes the equation of state. In this latter limit, since the fluid becomes incompressible, the sound-propagation velocity becomes infinite.

Of course, it is possible to start with kinetic theory and some assumptions about the nature of  the microscopic particles (eg., that they are monatomic and therefore there are no vibrational degrees of freedom) to derive the fluid dynamical equations from the Boltzmann transport equation\footnote{Typically this is done for non-relativistic set-ups, but a relativistic version is conceptually not very different.} \cite{Huang}.  The various non-relativistic limits can then be systematically obtained via different scaling limits of the particle velocities as the speed of light is taken to infinity. But unfortunately, this strategy is of limited use for conformal fluids because kinetic theory assumes a particle description, which is necessarily weakly coupled, and not immediately useful for the strongly coupled (near-)conformal fluids at RHIC and elsewhere. But it might still be interesting and fairly straightforward to see such a construction for weakly coupled conformal fluids, in particular for theories like ${\cal N}=4$ SYM which have marginal directions described by a tunable coupling.

\section*{Acknowledgments}

We would like to thank Daniel Arean, Christopher Eling (and through him, Yaron Oz), Bjarke Gud\hspace{-.21cm}${}^-$nason, Matt Kleban, Shiraz Minwalla, Mukund Rangamani, Bogdan Teaca and Ho-Ung Yee for discussions, comments and/or correspondence about fluids, gravity and/or fluid-gravity.

\appendix

\section*{{\bf Appendix A} \ On Non-Relativistic Limits of Barotropic Fluids}
\addcontentsline{toc}{section}{{\bf Appendix A} \ On Non-Relativistic Limits of Barotropic Fluids}
\renewcommand{\theequation}{A.\arabic{equation}}
\setcounter{equation}{0}

In this appendix, we discuss taking non-relativistic limits of compressible relativistic fluids. This is of some relevance to our work because our vortices become non-relativistic in the asymptotic region. Also, we want to emphasize certain distinctions between what has sometimes been called a non-relativistic limit in the literature and a scaling limit.  We will work with  ideal relativistic fluids\footnote{The non-relativistic limits of the kind we discuss here seem unexplored also in the case of dissipative fluids. It would be interesting to see how the ideas generalize.} whose energy-momentum tensor we repeat here in a more convenient form:
\bea
T^{\mu \nu}=\left(\frac{P}{c^2}+\rho\right)u^{\mu}u^{\nu}+\frac{P}{c^2}g^{\mu\nu}.\label{stressT}
\eea
The flow is described by its conservation.  The dimensionality of the spacetime is not crucial, we will take it to be 3+1, with metric signature is $\{-,+,+,+\}$, so $u^{\mu}u_{\mu}=-1$. We have included a factor of $c^2$ in the definition of $P$ because this is more natural in the non-relativistic limit. 
The precise definitions of $P$ and $\rho$ are meant so that they reproduce conventional definitions of pressure and temperature in the $c \rightarrow \infty$ limit. We will deal only with barotropic fluids, which are fluids whose equation of state is of the form $P=P(\rho)$. The conservation equation
\bea
\partial_{\mu}T^{\mu\nu}=0 \label{covdiv}
\eea
gives us four equations. The fluid is described by $P, \rho$ and the four components $u^{\mu}$. Together with the identity $u^{\mu}u_{\mu}=-1$, and the equation of state, this is enough to fully determine the fluid.\footnote{In general a fluid equation of state can include other thermodynamical state variables: it can include for instance, $T$. This addition of an extra variable will mean that we will need an extra equation to determine the fluid flow completely. This extra equation is usually a thermodynamical relation between the state variables. See \cite{Landau}.}

Ideal non-relativistic fluids are described by the continuity equation and the Euler equation. In the barotropic fluid case, the equations take the form \cite{Landau}:
\bea
\frac{\partial \rho}{\partial t}+{\bf \nabla}\cdot(\rho \ {\bf v})=0 \ \ \rm{(Continuity \ equation)} \label{NRcont}\\
\frac{\partial {\bf v}}{\partial t}+({\bf v}\cdot{\bf \nabla}) {\bf v}=-\frac{{\bf \nabla}P}{\rho}. \ \ \rm{(Euler's \ equation)} \label{NReuler}
\eea
These give rise to four equations and we also have the equation of state $P=P(\rho)$. Counting the three components of ${\bf v}$, the pressure $P$ and density $\rho$ we again have equal number of unknowns and equations. Even though we use the same symbols for the pressure and density in relativistic and non-relativistic fluids, they are {\em a-priori} logically distinct entities. The precise relation will be clarified when we discuss the precise way in which the non-relativistic limit is taken.

Our aim is to formalize the transition from (\ref{covdiv}) to (\ref{NRcont}-\ref{NReuler}). One approach for doing this for the specific case of conformal fluids was taken in \cite{Oz1} to relate conformal relativistic fluids to the incompressible Navier-Stokes equation. We will follow a somewhat different approach. It seems quite likely that what we present here was well-known to the ancients, but we will present it here because our perspective might be slightly different.

For the stress tensor given in (\ref{stressT}), the fluid-flow equations (\ref{covdiv}) can be transformed to the compact form
\bea
{\cal D} \rho + \left(\rho + \frac{P}{c^2}\right)\partial_\mu u^{\mu}=0, \label{covform1} \\
\frac{{\cal P}^{\mu\nu}\partial_\mu P}{c^2}+\left(\rho+\frac{P}{c^2}\right){\cal D}u^{\nu}=0. \label{covform2}
\eea
We have defined
\bea
{\cal P}^{\mu\nu}=g^{\mu\nu}+u^{\mu}u^{\nu}, \ \ {\cal D}=u^\mu \partial_\mu.
\eea
There is some redundancy in this description, as can be seen by contracting the second equation with $u_\mu$ and noticing that it yields an identity.

To make the non-relativistic limit intuitive, we will first write the relativistic equations in non-relativistic notation. To this end, we can identify $u^\mu=\gamma(1,{\bf v}/c)$, where $\gamma=1/\sqrt{1-v^2/c^2}$, which differs from the $\gamma$ defined in the text. This leads to the following useful explicit expressions:
\bea
\partial_\mu u^{\mu}=\frac{\gamma^3}{2c^3}Dv^2+\frac{\gamma}{c}{\bf \nabla}\cdot{\bf v}, \ \ \ {\cal D}=\frac{\gamma}{c}D, \\
{\cal P}^{\mu\nu}=\left(
\begin{array}{cc}
-1+\gamma^2 & \frac{\gamma^2 v_i}{c} \\
\frac{\gamma^2 v_j}{c} & \delta_{ij}+\frac{\gamma^2v_iv_j}{c^2}
\end{array}
\right).
\eea
where $D$ is the convective derivative $\partial_t+({\bf v}.{\bf \nabla})$ that appears in the Euler equation. Using these expressions and suitable linear combinations of (\ref{covform1}) and (\ref{covform2}), after some work, we can write down the relativistic fluid equations as
\bea
\partial_t \rho + {\bf \nabla}\cdot(\rho{\bf v})+\frac{\gamma^2}{2c^2}\left(\rho+\frac{P}{c^2}\right)D\ v^2+\frac{P {\bf \nabla}\cdot{\bf v}}{c^2}=0, \label{Rcont}\\
\frac{\partial {\bf v}}{\partial t}+({\bf v}\cdot{\bf \nabla}) {\bf v}=-\frac{1}{\gamma^2}\left(\frac{{\bf \nabla}P}{\rho+P/c^2}+\frac{{\bf v}}{c}\frac{\partial_t P}{(\rho+P/c^2)}\right). \label{Reuler}
\eea
The advantage of this particular form is that the structure of these two equations is an immediate generalization of the non-relativistic continuity (\ref{NRcont}) and Euler (\ref{NReuler}) equations. Now it is easy to see what we need to do in order to reproduce those equations from the relativistic ones. The appropriate limit is one where ${\bf v}, \ P$ and $\rho$ and their derivatives are held fixed, while we send $c \rightarrow \infty$. The equations immediately reduce to (\ref{NRcont}) and (\ref{NReuler}).

We see that the non-relativistic limit allows us to keep $P$ and $\rho$ independent, so the speed of sound $c_s=\sqrt{dP/d\rho}$ is held fixed as $c \rightarrow \infty$. This is essential, since otherwise the fluid would always end up being incompressible in the non-relativistic limit \cite{Oz1}. Note also that the $P$ that appears in the {\em incompressible} non-relativistic fluid equations (see eg., equation (17) in \cite{Oz1}) is in fact $P/\rho$ in our (non-relativistic) notation. The non-relativistic limit taken in \cite{Oz1} for conformal relativistic fluids, $P=\rho/3$, was a different scaling limit, which involved retaining other combinations (that include $c$) of quantities fixed, while sending $c \rightarrow \infty$. We think that the non-relativistic limit presented here is another very natural scaling limit. 

An interesting possibility is to see how the symmetries of the system change when we take our non-relativistic limit. Note that as a special case of the general thing we did above, we can consider the conformal case ($p=\rho/3$). The result is actually a non-relativistic fluid with $p=\rho/3$. This is {\em not} incompressible.  It would be interesting to see what the conformal invariance of the parent fluid becomes in this limit. For discussions on this issue, see \cite{MinNR, Oz1, Gopakumar, Horvathy}.

\section*{ {\bf Appendix B} \ The First Order System}
\addcontentsline{toc}{section}{{\bf Appendix B} \ The First Order System}
\renewcommand{\theequation}{B.\arabic{equation}}
\setcounter{equation}{0}

In this appendix we will re-write equations (\ref{alpha}-\ref{cale}) as a simple first order system. In principle, we can write $P$ and $\eta$ in (\ref{alpha}) in terms of the $T$ using (\ref{P1}-\ref{P2})\footnote{For a generic conformal fluid, the other thermodynamic variables are determined by temperature only up to proportionality constants. We will leave these constants arbitrary. 
}, and then solve the constraint for $T$ explicitly in terms of the fluid velocities and use it in the remaining differential equations. We have done this, and found that the results match in this approach with the approach presented below in some specific examples, as a check of our numerics. But in general this direct approach is cumbersome because the constraint is cubic and the resulting solution is very complicated, and the system is of mixed differential order.

Instead we will work with the linear (in derivatives) system where the expressions are in fact small enough to be presentable. First, we write $P$ and $\eta$ in terms of $T$ as before. Now, observe that using the definition of $A$, (\ref{alpha}) can be re-arranged to the form
\bea
g'=X_1(f, g, f', T, r). \label{app21}
\eea
(We won't write explicit forms until the very last step.). Note that (\ref{calc}) is just the definition of ${\cal C}$. Combining (\ref{calc}) and (\ref{cald}), we find that the second order differential equation that one ends up with is of the form
\bea
g''=X_2(f, g, f', g', T, T', r) \label{app22}
\eea
This expression is linear in $T'$. The third and final equation is (\ref{cale}), which can be written in the form
\bea
f'=F(f, g, T, r). \label{app23}
\eea
Taking a derivative of (\ref{app21}) we find that the derivative of $X_1$ should equal the right hand side of (\ref{app22}). This gives us a solution for $f''$ of the form
\bea
f''=X_3(f, g, f', g', T, T', r).
\eea
The right hand side of this in turn can be equated to the derivative of the right hand side of (\ref{app23}). The resulting equation can be solved for $T'$, and we find
\bea
T'=X_4(f, g, f', g', T, r).
\eea
At this stage, we can take the independent equations to be
\bea
f'=F(f, g, T, r), \ \ \ g'=X_1(f, g, f', T, r), \ \ \ T'=X_4(f, g, f', g', T, r).
\eea
Using the first of these to substitute for $f'$ in the RHS of the second, and then using the resulting equation as well in the third, we can bring it to the final form
\bea
f'=F(f, g, T, r), \ \ g'=G(f, g, T, r), \ \ \ T'=H(f, g, T, r),
\eea
which is the promised first order system. The explicit forms of $F, G$ and $H$ are
\bea
F&=&\frac{c_2+2 b r^4 f^3 T^2-2 r^4 f g T^2 (b g-3 a T)}{2 r^3 \left(b+b r^2 g^2\right) T^2
 }, \nonumber \\
G&=&-\frac{\left( c_2 f +2 c_1 \gamma - 4 b r^4 f^2 g^2 T^2-6 a r^2 g T^3-6 a r^4 g^3 T^3\right)}{2 b r^3 g T^2+2 b r^5 g^3 T^2}, \nonumber \\
H&=&\frac{\left\{\begin{array}{l}
4 c_1^2 {\gamma}+4 {c_1} {c_2} r^2 f^3+{c_2}^2 {\gamma} g^2-12 a c_1 r^2 g T^3-24 a b {\gamma} r^6 g^5 T^5+\\
\hspace{0.2in} -12 a r^4 g^3 T^3 \left(c_1+2 b {\gamma} T^2\right)+2 {c_2} f \left(2 {c_1}+2 {c_1} r^2g^2-3 a {\gamma} r^2 g T^3\right)+\\
\hspace{1in}  + f^2 \left({\gamma} \left({c_2}^2+4 {c_1}^2 r^2\right)-12 a {c_1} r^4 g T^3-24 a b{\gamma} r^6 g^3 T^5\right)
                \end{array}\right\}}{24 a b r^5 g^3 \left(1+r^2 g^2\right) T^4} \nonumber
\eea
Note that we have kept the transport coefficients general in these expressions. To go to the black brane values, which is what we use in the numerics, we can use (\ref{P2}). In the above equations $\gamma \equiv\sqrt{1+r^2 f^2+r^2 g^2}$ as before.

\end{document}